\documentclass[11pt]{article}

\newcommand{\phlam}{\ensuremath{\phi_\lambda}}
\newcommand{\phkap}{\ensuremath{\phi_\kappa}}

\newcommand{\phtri}{\ensuremath{\phi_{A_0}}}
\newcommand{\nn}{\nonumber}

\usepackage{calc}
\usepackage{rotating}
\usepackage[english]{babel}
\usepackage{graphicx}
\usepackage{subfig}
\usepackage{float}
\usepackage{amsmath}
\usepackage{amssymb}
\usepackage{amsthm}
\usepackage{latexsym}
\usepackage{dcolumn}
\usepackage{geometry}
\usepackage{color}
\usepackage{hyperref}
\usepackage{mciteplus}
\newcommand{\Vac}[1]{\bigg\langle{#1}\bigg\rangle}

\def\gtwid{\mathrel{\raise.3ex\hbox{$>$\kern-.75em\lower1ex\hbox{$\sim
$}}}}
\def\vio{\mathrel{\hbox{$E$\kern-.60em\hbox{$/
$}}}}
\begin{document}

\title{Novel Higgs-to-125\,GeV Higgs boson decays in the complex NMSSM}

\author{\bf Shoaib Munir\footnote{Present address:
 Department of Physics and Astronomy, Uppsala University, Box 516,
    SE-751\,20 Uppsala, Sweden.\newline{email: shoaib.munir@physics.uu.se}} \\
National Centre for Nuclear Research, Ho{\. z}a 69, 00-681 Warsaw, Poland}

\maketitle

 \begin{abstract} 
 In the Next-to-Minimal Supersymmetric Standard Model (NMSSM)
  a variety of parameter configurations yields a Higgs boson consistent
  with the one observed at the LHC. Additionally, the Higgs sector of
  the model can contain explicit CP-violating phases even at the
  tree level, in contrast with the Minimal Supersymmetric Standard
  Model (MSSM). In this article we present the one-loop
  Higgs boson mass matrix of the complex NMSSM in the
  renormalisation-group-improved effective
  potential approach. We also present the trilinear Higgs boson
  self-couplings as well as various partial decay widths of a generic
  CP-mixed Higgs boson in the model. We then analyse a very
  interesting phenomenological
  scenario wherein the decay of a relatively light
  pseudoscalar-like Higgs boson into $\sim 125$\,GeV SM-like
  Higgs boson(s) is induced by non-zero CP-violating phases. We discuss in
  detail a few benchmark cases in which such a decay can contribute
  significantly to the production of SM-like Higgs bosons at
  the LHC on top of the gluon fusion process. It can thus be partially
  responsible for the $\gamma\gamma$ excess near 125\,GeV due 
  to the subsequent decay of the SM-like Higgs boson. Such a scenario is extremely difficult to realize in the complex MSSM and, if probed at the LHC, it could provide an indication of the non-minimal nature of supersymmetry.
 \end{abstract} 

\section{Introduction}\label{intro}

The new particle with mass around 125\,GeV first observed by the CMS and
ATLAS experimental collaborations at the Large Hadron Collider (LHC)
in July 2012\,\cite{Aad:2012tfa,Chatrchyan:2012ufa} seems
to be increasingly consistent with the Higgs boson of
the Standard Model (SM)\,\cite{Meola:2013xva,CMS-H-twiki,ATLAS-H-twiki}.
However, there is growing evidence from other collider experiments as
well as from astroparticle physics and cosmology that the SM fails to
provide a complete description of nature and that there must lie
physics beyond it. One of the most important yet unresolved issues in particle
physics is that of CP violation. Although it was first discovered
experimentally\,\cite{Christenson:1964fg} many decades ago, its only source in the SM\,\cite{Kobayashi:1973fv} does not prove sufficient
to explain the observed baryon asymmetry in the Universe. Therefore, a
variety of sources of CP violation beyond the SM have been proposed in the
literature (for a review, see\,\cite{Ibrahim:2007fb} and references therein), but these remain
hidden to this day. 

In models with supersymmetry (SUSY), the soft masses and couplings of the
superpartners of SM particles as well as the soft Higgs sector
parameters can very well be complex and can thus  
explain baryogenesis by generating the desired amount of CP-violation. 
The Higgs sector of the Minimal Supersymmetric Standard
Model (MSSM) does not contain CP-violating (CPV) phases at the tree
level and these are only induced at the one-loop level by the sfermion
sector\,\cite{Pilaftsis:1998pe,*Pilaftsis:1998dd,Pilaftsis:1999qt,*Carena:2000yi,Choi:2000wz,*Carena:2001fw,*Carena:2002bb,*Choi:2004kq,*Frank:2006yh,*Heinemeyer:2007aq}. These phases can substantially modify
both the mass spectrum and production/decay rates of the Higgs
bosons\,\cite{Demir:1999hj,*Kane:2000aq,*Carena:2000ks,*Arhrib:2001pg,*Choi:2001pg,*Choi:2002zp,*Ellis:2005ika,*Williams:2011bu,*Fritzsche:2011nr}
and can at the same time provide a solution
to electroweak baryogenesis\,\cite{Carena:1996wj,*Carena:1997gx,*Carena:2000id,*Carena:2002ss,*Carena:2008vj,*Cirigliano:2009yd,*Carena:2012np}. However, these phases are also strongly constrained by the measurements of
fermionic electric dipole moments (EDMs)\,\cite{Baker:2006ts,*Commins:2007zz,*Griffith:2009zz}. In the
context of the LHC, the impact
of the CPV phases on the phenomenology of the MSSM Higgs bosons
was studied in detail in\,\cite{Dedes:1999sj,*Dedes:1999zh,*Choi:2001iu,*Ellis:2004fs,*Moretti:2007th,*Hesselbach:2007en,*Hesselbach:2009gw} prior to the
Higgs boson discovery and has been revisited
in\,\cite{Chakraborty:2013si} afterwards.
	
In the Next-to-Minimal Supersymmetric
Standard Model (NMSSM)\,\cite{Fayet:1974pd,*Ellis:1988er,Durand:1988rg,*Drees:1988fc,Miller:2003ay} (see, e.g., \cite{Ellwanger:2009dp,Maniatis:2009re} for reviews) the
presence of an additional Higgs singlet field besides the
two MSSM doublets has some very interesting phenomenological
implications. In this model either of the two lightest CP-even Higgs
bosons, $h_1$ and $h_2$, can play the role of the observed SM-like
Higgs boson with a mass around 125\,GeV\,\cite{Ellwanger:2011aa}. 
In fact in the NMSSM it is also possible to have $h_1$ and $h_2$
almost degenerate in mass
around 125\,GeV\,\cite{Gunion:2012gc}, so that the observed signal is
actually a superposition of two individual peaks due to each of these,
and likewise for $h_1$ and $a_1$, the lightest pseudoscalar of the
model\,\cite{Munir:2013wka}. Additionally, in some regions of the
NMSSM parameter space the singlet-like scalar or
 pseudoscalar of the model can be considerably lighter than the
SM-like Higgs boson. In these regions
the SM-like Higgs boson can decay via such `invisible' channels, causing
a significant suppression of the $\gamma\gamma$ and $ZZ$ signal rates, as
studied recently in\,\cite{Ananthanarayan:2013fga,Cao:2013gba}. 

The NMSSM contains some
new couplings in the Higgs sector which, if assumed to be complex, can
result in new CPV phases even at the tree level, conversely
to the MSSM. Indeed, additional MSSM-like phases also appear in the
Higgs boson mass matrix beyond the born
approximation. Non-zero CPV phases can substantially modify the phenomenology 
of the $\sim 125$\,GeV SM-like Higgs boson in the NMSSM, as studied
recently in\,\cite{Moretti:2013lya}.
But, like the MSSM, the measurements of fermionic EDMs can put
strong constraints on the allowed values of the CPV phases in the
NMSSM also. However, the
conditions under which these EDM constraints can be
avoided in the MSSM\,\cite{Pilaftsis:1999qt,*Carena:2000yi,Abel:2001vy} 
in fact also apply in this model. One can, for example, assign
very heavy soft masses to the sfermions of the first two generations in
order to  minimize their contribution to the EDMs. Alternatively,
one can argue that the phase
combinations occurring in the EDMs can be different from the ones
inducing Higgs boson mixing \cite{Haba:1996bg,*Ibrahim:1998je,*Boz:2005sf,*Ellis:2008zy,*Li:2010ax}. 

The complete one-loop Higgs mass
matrix has been derived in\,\cite{Graf:2012hh} in the Feynman
diagrammatic approach.
In the renormalisation-group (RG)-improved effective potential
approach the neutral Higgs sector of the complex NMSSM (cNMSSM) has previously
been studied in detail
in\,\cite{Ham:2001wt,*Funakubo:2004ka,Cheung:2010ba,*Cheung:2011wn}, 
including only the dominant one-loop corrections
from the (s)quark and gauge sectors.
In this article, we provide the RG-improved one-loop Higgs mass matrix
of the cNMSSM in the effective
potential approach in which the complete set of dominant corrections
from the third generation (s)quark, stau, gauge as well as
chargino/neutralino sectors have been included. We also present the
tree level expression for the trilinear Higgs boson self-couplings in the
cNMSSM. These couplings are extremely important for studying the
LHC phenomenology of Higgs bosons in the model.  
Moreover, we present the set of expressions for
 partial decay widths of a CPV Higgs boson.

The Higgs boson mass matrix and decay widths provided here
have been implemented in a comprehensive fortran package for conveniently carrying out 
phenomenological studies of the cNMSSM Higgs sector. 
Using this package we analyse in this article a very interesting
scenario made possible
by non-zero CPV phases in the NMSSM, owing to the fact that the five
neutral Higgs bosons of the model no longer carry
definite CP assignments. The scalars and pseudoscalars of
the CP-conserving (CPC) limit thus couple to one
another, which implies that any of these Higgs bosons can have a non-zero decay
width into a pair of lighter ones, when kinematically
allowed. We argue that such a scenario can be of particular
importance in the context of the recent LHC discovery. The reason is
that it is very much probable for the lighter of the two
pseudoscalar-like Higgs bosons to have a mass $\sim 250$\,GeV, particularly when one of the
scalar-like Higgs bosons is required to have
SM-like $\gamma\gamma$ and $ZZ$ signal rates and a mass near
125\,GeV. Such a mass would result a much larger branching ratio (BR)
of this Higgs boson into a pair of the SM-like Higgs
bosons compared to that of the other, typically much heavier, scalar-like
Higgs bosons, despite a relatively much smaller trilinear coupling.  

However, despite having a large BR into lighter Higgs bosons, the
above mentioned $\sim 250$\,GeV boson 
can be very difficult to produce at the LHC on account of being 
singlet-like and thus
having a considerably reduced coupling to two gluons. Therefore, 
the relative probability of its production
in the gluon fusion mode also needs to be taken into account in the
above scenario. For this purpose, we define an auxiliary signal rate, similar
to the conventional `reduced cross section', which quantifies 
the contribution of the $\sim 250$\,GeV boson to the production of the 
SM-like Higgs bosons, decaying eventually into photons pairs, at the LHC. 
We then select representative points from three distinct regions in the
cNMSSM parameter space wherein
the $\sim 125$\,GeV SM-like Higgs boson is either $h_1$ or
$h_2$, the lightest and next-to-lightest of the five neutral Higgs bosons,
respectively, to investigate our scenario of interest. We discuss in
detail the impact of the variation in the most relevant of the CPV
phases on our auxiliary signal rate in each of these cases. We conclude that
for large values of the phase, this rate can become quite significant, 
reaching a few tens of percent of
the direct production rate of the SM-like Higgs boson in the gluon
fusion channel.   

The article is organised as follows. In the next section we will 
give details of the cNMSSM Higgs mass matrix at the tree
level and the one-loop as well as logarithmically
enhanced dominant two-loop corrections to it. In
Sect.\,\ref{sec:couplings} we will present the expressions for the
trilinear self-couplings of the Higgs bosons and will also define notation for their
couplings to other model particles. In
Sect.\,\ref{sec:decays} we will provide detailed expressions for all possible
two-body partial decay widths of the Higgs boson in the presence of
CPV phases. In Sect.\,\ref{sec:results}, after discussing at
length our scenario of interest, we will present our numerical results for the three points
investigated. We will summarise our findings in Sect.\,\ref{sec:summary}.

\section{\label{sec:mass-matrix} Higgs sector of the cNMSSM}

As noted in the introduction, the NMSSM contains a singlet Higgs
superfield, $\widehat{S}$, besides the two MSSM $SU(2)_L$ doublet
superfields,
\begin{eqnarray}
\widehat{H}_u = \left(\begin{array}{c} \widehat{H}_u^+ \\ \widehat{H}_u^0
\end{array}\right)\,,\;
\widehat{H}_d = \left(\begin{array}{c} \widehat{H}_d^0 \\ \widehat{H}_d^-
\end{array}\right)\,.
\end{eqnarray}
The scale-invariant superpotential of the cNMSSM is thus written as
\begin{equation}
\label{eq:superpot}
W_{\rm NMSSM}\ =\ {\rm MSSM\;Yukawa\;terms} \: +\: 
\lambda \widehat{S} \widehat{H}_u \widehat{H}_d \: + \:
\frac{\kappa}{3}\ \widehat{S}^3\,,
\end{equation}
where $\lambda\equiv |\lambda|e^{i\phi_\lambda}$ and $\kappa\equiv
|\kappa|e^{i\phi_\kappa}$ are dimensionless complex Yukawa couplings. The
second term in the above superpotential 
replaces the Higgs-higgsino mass term, $\mu \widehat{H}_u \widehat{H}_d$,
of the MSSM superpotential and the last cubic term explicitly breaks the dangerous
$U(1)_{PQ}$ symmetry, introducing in turn a discrete $Z_3$
symmetry. Upon breaking the electroweak symmetry, the singlet field
acquires a vacuum expectation value (VeV), $s$, naturally of the order of the
SUSY-breaking scale, $M_{\rm SUSY}$, and an effective $\mu$-term, $\mu_{\rm eff} = \lambda s$, is generated.

\subsection{Tree level Higgs potential and mass matrix}

The superpotential in Eq.\,(\ref{eq:superpot}) leads
to the tree level Higgs potential containing the $D$-, $F$-
and soft SUSY-breaking terms:
\begin{eqnarray}
\label{eq:Higgspot}
V_0 & = & \left|\lambda \left(H_u^+ H_d^- - H_u^0
H_d^0\right) + \kappa S^2 \right|^2 \nn \\
&&+\left(m_{H_u}^2 + \left|\mu + \lambda S\right|^2\right) 
\left(\left|H_u^0\right|^2 + \left|H_u^+\right|^2\right) 
+\left(m_{H_d}^2 + \left|\mu + \lambda S\right|^2\right) 
\left(\left|H_d^0\right|^2 + \left|H_d^-\right|^2\right) \nn \\
&&+\frac{g^2}{4}\left(\left|H_u^0\right|^2 +
\left|H_u^+\right|^2 - \left|H_d^0\right|^2 -
\left|H_d^-\right|^2\right)^2
+\frac{g_2^2}{2}\left|H_u^+ H_d^{0*} + H_u^0 H_d^{-*}\right|^2\nn \\
&&+m_{S}^2 |S|^2
+\big( \lambda A_\lambda \left(H_u^+ H_d^- - H_u^0 H_d^0\right) S + 
\frac{1}{3} \kappa A_\kappa\, S^3  + \mathrm{h.c.}\big)\,,
\end{eqnarray}
where $g^2\equiv \frac{g_1^2 + g_2^2}{2}$, with $g_1$ and $g_2$ being
the $U(1)_Y$ and $SU(2)_L$ gauge couplings, respectively, and $A_\lambda \equiv |A_\lambda|e^{i\phi_{A_\lambda}}$
and $A_\kappa \equiv |A_\kappa|e^{i\phi_{A_\kappa}}$ are dimensionful soft SUSY-breaking
trilinear couplings. These, along with $\lambda$ and $\kappa$, are the
only complex parameters appearing in the tree level Higgs potential,
since the soft SUSY-breaking masses $m_{H_u}^2$, $m_{H_d}^2$ and $m_{H_u}^2$ are real.

In order to obtain the physical Higgs states, the above potential is
expanded around the VeVs of the three Higgs fields as
\begin{eqnarray}
H_d^0&=&
\hphantom{e^{i\theta}}\,
\left(
\begin{array}{c}
\frac{1}{\sqrt{2}}\,(v_d+H_{dR}+iH_{dI}) \\
H_d^-
\end{array}
\right)\,, \nonumber \\[1mm]
H_u^0&=&
e^{i\theta}\,\left(
\begin{array}{c}
H_u^+\\
\frac{1}{\sqrt{2}}\,(v_u+H_{uR}+i H_{uI})
\end{array}
\right)\,, \nonumber \\[1mm]
S&=&\frac{e^{i\varphi}}{\sqrt{2}}\,(s+S_R+iS_I)\,.
\label{eq:higgsparam}
\end{eqnarray}
The potential in Eq.\,(\ref{eq:Higgspot}) then has a minimum at non-vanishing $v_u$,
$v_d$ and $s$ only if the following so-called tadpole conditions are satisfied:
\begin{eqnarray}
\frac{1}{v_d}\Vac{\frac{\partial V_0}{\partial H_{dR}}}
&=&m_{H_d}^{2}+\frac{g^2}{4}(v_{d}^{2}-v_{u}^{2})-R_\lambda\frac{v_us}{v_d}
        +\frac{|\lambda|^2}{2}(v_u^2+s^2)-\frac{1}{2}\mathcal{R}\frac{v_u
          s^2}{v_d}=0\,,
\nonumber \\
\frac{1}{v_u}\Vac{\frac{\partial V_0}{\partial H_{uR}}}
&=&m_{H_u}^{2}-\frac{g^2}{4}(v_{d}^{2}-v_{u}^{2})-R_\lambda\frac{v_ds}{v_u}
        +\frac{|\lambda|^2}{2}(v_d^2+s^2)-\frac{1}{2}\mathcal{R}\frac{v_d
          s^2}{v_u}=0\,,
\nonumber \\
\frac{1}{s}\Vac{\frac{\partial V_0}{\partial S_R}}
&=&m_S^2-R_\lambda\frac{v_dv_u}{s}+\frac{|\lambda|^2}{2}(v_d^2+v_u^2)
        +|\kappa|^2s^2-\mathcal{R}v_dv_u+R_\kappa s=0\,,
\label{eq:cpeventad} \\
\frac{1}{v_u}\Vac{\frac{\partial V_0}{\partial H_{dI}}}
&=&\frac{1}{v_d}\Vac{\frac{\partial V_0}{\partial H_{uI}}}
        =I_\lambda s+\frac{1}{2}\mathcal{I}s^2=0\,,
\nonumber \\
\frac{1}{s}\Vac{\frac{\partial V_0}{\partial S_I}}
&=&I_\lambda \frac{v_dv_u}{s}-\mathcal{I}v_dv_u-I_\kappa s=0\,,
\label{eq:cpoddtad}
\end{eqnarray}
where we have defined
\begin{eqnarray}
\label{eq:pardef}
\mathcal{R} &=& |\lambda| |\kappa|\, \cos(\phi^\prime_\lambda-\phi^\prime_\kappa)\,,
\hspace{1.0cm}
\mathcal{I} = |\lambda| |\kappa|\, \sin(\phi^\prime_\lambda-\phi^\prime_\kappa)\,,
\nonumber \\
R_\lambda &=& \frac{|\lambda| |A_\lambda|}{\sqrt{2}}\,
\cos(\phi^\prime_\lambda+\phi_{A_\lambda})\,,
\ \ \
R_\kappa = \frac{|\kappa| |A_\kappa|}{\sqrt{2}}\,
\cos(\phi^\prime_\kappa+\phi_{A_\kappa})\,, \nonumber\\
I_\lambda &=& \frac{|\lambda| |A_\lambda|}{\sqrt{2}}\,
\sin(\phi^\prime_\lambda+\phi_{A_\lambda}) \,,
\ \ \
I_\kappa = \frac{|\kappa| |A_\kappa|}{\sqrt{2}}\,
\sin(\phi^\prime_\kappa+\phi_{A_\kappa})  \,,
\end{eqnarray}
with
\begin{equation}
\phi^\prime_\lambda \equiv \phi_\lambda+\theta+\varphi \ \ \ {\rm and} \ \ \
\phi^\prime_\kappa \equiv \phi_\kappa+3\varphi\,.
\end{equation}
The parameters $I_\lambda$ and $I_\kappa$ can be re-expressed in terms of $\mathcal{I}$
using the CP-odd tadpole conditions in Eq.\,(\ref{eq:cpoddtad}) as
\begin{eqnarray}
I_\lambda = -
\frac{1}{2}\,\mathcal{I}\, s\,,
\hspace{1.0cm}
I_\kappa = -\frac{3}{2}\,\mathcal{I}\, \frac{v_dv_u}{s}\,.
\label{eq:ilik}
\end{eqnarray}
Then the phase combinations 
 $\phi^\prime_\lambda+\phi_{A_\lambda}$ and
 $\phi^\prime_\kappa+\phi_{A_\kappa}$ are determined up to a 
twofold ambiguity by $\phi^\prime_\lambda-\phi^\prime_\kappa$, which
is thus the only remaining physical CP phase 
at the tree level.
The three CP-even tadpole conditions in Eq.\,(\ref{eq:cpeventad}), on
the other hand, can be used to remove the soft mass parameters $m_{H_u}^2$, $m_{H_d}^2$ and $m_S^2$.

The $6\times 6$ neutral Higgs mass matrix, obtained by taking the second derivative of
 the potential in Eq.~(\ref{eq:Higgspot}) evaluated at the vacuum, can be cast into
 the form
\begin{eqnarray}
\label{eq:mhiggs}
\hspace{-1.0cm}
\mathcal{M}^2_0 =
\left(
        \begin{array}{cc}
\mathcal{M}_S^2 & \mathcal{M}^{2}_{SP} \\
(\mathcal{M}^{2}_{SP})^T & \mathcal{M}_P^{2}  
                \end{array}
\right)\,,
\end{eqnarray}
in the basis ${\bf H}^T=(H_{dR},\,H_{uR},\,S_R,\,H_{dI},\,H_{uI},\,S_I)$. The elements of the top left $3\times 3$
CP-even block in the above equation are given as
\begin{eqnarray}
{\cal M}_{S,11}^2&=&\frac{g^2}{2}v_d^2(Q)
        +\left(R_\lambda+\frac{\mathcal{R}s(Q)}{2}\right)s(Q)\tan\beta\,, \nonumber \\
{\cal M}_{S,22}^2&=&\frac{g^2}{2}v_u^2(Q)
        +\left(R_\lambda+\frac{\mathcal{R}s(Q)}{2}\right)\frac{s(Q)}{\tan\beta}\,, \nonumber \\
{\cal M}_{S,33}^2&=&R_\lambda\frac{v_d(Q)v_u(Q)}{s(Q)}+2|\kappa|^2s(Q)^2+R_\kappa
s(Q) \,, \nonumber \\
{\cal M}_{S,12}^2&=&({\cal M}_{S,21}^2)
        =\left(-\frac{g_1^2+g_2^2}{4}+|\lambda|^2\right)v_d(Q)v_u(Q)
        -\left(R_\lambda+\frac{\mathcal{R}s(Q)}{2}\right)s(Q) \,, \nonumber \\
{\cal M}_{S,13}^2&=&({\cal M}_{S,31}^2)
        =-R_\lambda v_u(Q)+|\lambda|^2v_d(Q)s(Q)-\mathcal{R}v_u(Q)s(Q) \,,\nonumber \\
{\cal M}_{S,23}^2&=&({\cal M}_{S,32}^2)
        =-R_\lambda v_d(Q)+|\lambda|^2v_u(Q)s(Q)-\mathcal{R}v_d(Q)s(Q) \,,
\end{eqnarray}
where $v_u(Q)$, $v_d(Q)$ and $s(Q)$ are the three Higgs VeVs
defined at the scale $Q^2= M_{\rm
SUSY}^2$ and $\tan\beta\equiv v_u(Q)/v_d(Q)$. The bottom right CP-odd block in Eq.\,(\ref{eq:mhiggs}) is given as
\begin{eqnarray}
 {\cal M}_{P,11}^2&=& \left(R_\lambda+\frac{\mathcal{R}s(Q)}{2}\right)s(Q)\tan\beta\,, \nonumber \\
 {\cal M}_{P,22}^2&=& \left(R_\lambda+\frac{\mathcal{R}s(Q)}{2}\right)\frac{s(Q)}{\tan\beta}\,,
 \nonumber \\
  {\cal M}_{P,33}^2&=& R_\lambda\frac{v_d(Q)v_u(Q)}{s(Q)}+2\mathcal{R}v_d(Q)v_u(Q) -3R_\kappa s(Q)\,,\nonumber \\
 {\cal M}_{P,12}^2&=&({\cal M}_{P,21}^2) =
 \left(R_\lambda+\frac{\mathcal{R}s(Q)}{2}\right)s(Q) \,,\nonumber \\
 {\cal M}_{S,13}^2&=&({\cal M}_{S,31}^2) =
 \big(R_\lambda-\mathcal{R}s(Q)\big)v_u(Q) \,,\nonumber \\
{\cal M}_{S,23}^2&=&({\cal M}_{S,32}^2) =
 \big(R_\lambda-\mathcal{R}s(Q)\big)v_d(Q)\,, 
\end{eqnarray}
and the off-diagonal CP-mixing block reads
\begin{equation}
 \mathcal{M}_{SP}^{2}=\left(\begin{array}{ccc}
 0& 0&
 -\frac{3}{2}\mathcal{I}sv_u\\
 0& 0&
 -\frac{3}{2}\mathcal{I}sv_d\\
 \frac{1}{2}\mathcal{I}sv_u&
 \frac{1}{2}\mathcal{I}sv_d&
 2\mathcal{I}v_dv_u
\end{array}\right)\,.
\end{equation}
\subsection{\label{sec:potential}RG-improved one-loop effective potential}

The one-loop corrections to the effective potential are given by the
Coleman-Weinberg formula (in the $\overline{\text{DR}}$ scheme with an
ultraviolet cutoff $M_{\rm SUSY}^2$) as
\begin{eqnarray}
\label{c.13e}
\Delta V_{\rm eff} = \frac{1}{64\pi^2} {\text{STr}}\, M^4 
\left[ \ln\left(\frac{M^2}{M_{\rm SUSY}^2}\right) -\frac{3}{2} \right]\,.
\end{eqnarray}
As a result of these corrections the Higgs mass matrix gets
modified so that
\begin{eqnarray}
\label{eq:mhiggs-nor}
\mathcal{M}^2_H = \mathcal{M}^2_0 + \Delta\mathcal{M}^2_{\rm eff}\,.
\end{eqnarray}
In the following we present analytical expressions for the
corrections $\Delta\mathcal{M}^2_{\rm eff}$ above. 
These corrections have been adopted from \cite{Ellwanger:2009dp}
and modified to explicitly include the CPV phases. They are thus of
the same order as those implemented in the publicly available package NMSSMTools-v3.2.4 \cite{NMSSMTools}. 

\subsubsection{Top and bottom squark contributions} 
 
Some of the radiative corrections due to the stop and
sbottom loops can be accounted for by the following shift in the Higgs mass matrix:
\begin{eqnarray}
\label{c.17e}
A_\lambda \to  A'_\lambda =
A_\lambda + \frac{3h_t^2}{16\pi^2}A_t\,
f_t + \frac{3h_b^2}{16\pi^2}A_b\, f_b\,,
\end{eqnarray}
where $h_t \equiv \frac{2 m_t}{v_u}$ and $h_b \equiv \frac{2
  m_b}{v_d}$ are the Yukawa couplings of top and bottom
quarks, with $m_t$ and $m_b$ being their respective masses. Note that
these Yukawa couplings have complex phases in general. However, we
assume them to be real, since their non-zero phases
can always be reabsorbed by redefining the quark
fields\,\cite{Kobayashi:1973fv} when generation mixing is neglected,
which is the case here. 
$A_t\equiv |A_t|e^{i\phi_{A_t}}$ and $A_b\equiv |A_b|e^{i\phi_{A_b}}$
are the complex soft SUSY-breaking counterparts of these Yukawa
couplings for the top and bottom squarks, respectively.

The above shift results 
in the redefinition of the parameters $\mathcal{R}$ and
$\mathcal{I}$ given in Eq.\,(\ref{eq:pardef}) and a subsequent 
improvement in the relation between the latter and $I_\lambda$ given in Eq.\,(\ref{eq:ilik}).
It also takes care of the $\sim h^4_{
t,b}$ radiative corrections to $\mathcal{M}^2_P$. The remaining corrections $\sim h_t^2
\equiv h_t^2(M_{\rm SUSY}^2)$ and $\sim h_b^2 \equiv h_b^2(M_{\rm SUSY}^2)$ to $\mathcal{M}_S^2$ are written as
\begin{eqnarray}
{\Delta \cal M}_{S,11}^2 & = & 
\frac{3 h_b^2m_b^2}{8\pi^2}\Big(
-|A_b|^2B'_b\,g_b
+ 2 |A_b| B_{b}L_{\tilde{b}} + L_{\tilde{b}b} \Big) 
- \frac{3h_t^2
m_t^2}{8\pi^2}|\mu|^2B'_t\,g_t\,, \nn\\
{\Delta \cal M}_{S,22}^2 & = & 
\frac{3 h_t^2m_t^2}{8\pi^2}\Big(
-|A_t|^2B'_t\,g_t
+ 2 |A_t| B_{t}L_{\tilde{t}} + L_{\tilde{t}t} \Big) 
- \frac{3h_b^2
m_b^2}{8\pi^2}|\mu|^2B'_b\,g_b\,, \nn\\
{\Delta \cal M}_{S,33}^2 & = & 
- \frac{3 h_t^2m_t^2}{16\pi^2}|\lambda|^2v_d^2(Q)B'_t\,g_t
- \frac{3 h_b^2m_b^2}{16\pi^2}|\lambda|^2v_u^2(Q)B'_b\,g_b
\,, \nn\\
{\Delta \cal M}_{S,12}^2 & = & 
\frac{3 h_t^2m_t^2}{8\pi^2}|\mu|\Big(|A_t|B'_t\,g_t \cos(\phlam' +
  \phi_{A_t}) -
  \frac{|A_t|\cos(\phlam'+\phi_{A_t})
  +|\mu|\cot\beta}{m_{\tilde{t}_2^2}-m_{\tilde{t}_1^2}}\Big) \nn\\ 
&&+\frac{3 h_b^2m_b^2}{8\pi^2}|\mu|\Big(|A_b|B'_b\,g_b\cos(\phlam' +
  \phi_{A_b}) -
  \frac{|A_b|\cos(\phlam'+\phi_{A_b})
  +|\mu|\tan\beta}{m_{\tilde{b}_2^2}-m_{\tilde{b}_1^2}}\Big) \,, \nn\\
{\Delta \cal M}_{S,13}^2 & = & 
\frac{3 h_b^2m_b^2|\lambda|v_u(Q)}{8 \sqrt{2}\pi^2}\Big(|A_b|B'_b\,g_b\cos(\phlam' +
  \phi_{A_b}) -
  \frac{|A_b|\cos(\phlam'+\phi_{A_b})
  +|\mu|\tan\beta}{m_{\tilde{b}_2^2}-m_{\tilde{b}_1^2}}\Big)\nn \\
&& +\frac{3 h_t^2}{ 64\pi^2}|\lambda|^2 s(Q) \, v_d(Q)(4\, f_t-4m_t^2B'_t\,g_t) 
\,,\nn
\end{eqnarray}
\begin{eqnarray} 
{\Delta \cal M}_{S,23}^2 & = & 
\frac{3 h_t^2m_t^2|\lambda|v_d(Q)}{8 \sqrt{2}\pi^2}\Big(|A_t|B'_t\,g_t\cos(\phlam' +
  \phi_{A_t}) -
  \frac{|A_t|\cos(\phlam'+\phi_{A_t})
  +|\mu|\cot\beta}{m_{\tilde{t}_2^2}-m_{\tilde{t}_1^2}}\Big)\nn \\
&& +\frac{3 h_b^2}{ 64\pi^2}|\lambda|^2 s(Q) \, v_u(Q)(4\, f_b-4m_b^2B'_b\,g_b) 
\,,
\label{c.18e}
\end{eqnarray}
where $|\mu| \equiv
|\mu_{\rm eff}|/\sqrt{2} = |\lambda| s(Q)/\sqrt{2}$, $m_t$ and $m_b$
are the masses of $t$ and $b$ quarks, respectively, and the squark
masses $m_{\tilde q}$ are given in
Appendix\,\ref{sec:app-A}. Also in the above equations
\begin{eqnarray}
B_{t} = \frac{|A_t|-
  |\mu|\cot\beta\cos(\phlam'+\phi_{A_t})}{m_{\tilde{t}_2^2}-m_{\tilde{t}_1^2}}\,,\ \ \
B_{b} = \frac{|A_b|-
  |\mu|\tan\beta\cos(\phlam'+\phi_{A_b})}{m_{\tilde{b}_2^2}-m_{\tilde{b}_1^2}}\,,\nn 
\end{eqnarray}
\vspace*{-0.5cm}\begin{eqnarray}
B'_t &=& \frac{|A_t|^2 + |\mu|^2 \cot^2\beta -
  2|\mu||A_t|\cot\beta\cos(\phlam'-\phi_{A_t})}{(m_{\tilde{t}_2^2}-m_{\tilde{t}_1^2})^2}\,,\nn \\
B'_b &=& \frac{|A_b|^2 + |\mu|^2 \tan^2\beta -
  2|\mu||A_b|\tan\beta\cos(\phlam'-\phi_{A_b})}{(m_{\tilde{b}_2^2}-m_{\tilde{b}_1^2})^2}\,,
\end{eqnarray}
 and the quantities $L_{\tilde{f}}$, $L_{\tilde{f}f}$, $f_f$ and
 $g_f$ are given in Appendix\,\ref{sec:app-B}. $\mathcal{M}_{SP}^{2}$ also receives the corresponding corrections given as 
\begin{eqnarray}
{\Delta \cal M}_{SP,11}^2 & = & 
-\frac{3 h_b^2m_b^2}{4\pi^2}
 \frac{
  |A_b||\mu|\tan\beta\sin(\phlam'+\phi_{A_b})}{m_{\tilde{b}_2^2}-m_{\tilde{b}_1^2}}L_{\tilde{b}}  \,, \nn\\
{\Delta \cal M}_{SP,22}^2 & = & 
-\frac{3 h_t^2m_t^2}{4\pi^2}
\frac{|A_t| 
  |\mu|\cot\beta\sin(\phlam'+\phi_{A_t})}{m_{\tilde{t}_2^2}-m_{\tilde{t}_1^2}}L_{\tilde{t}} \,, \nn\\
{\Delta \cal M}_{SP,12}^2 & = & 
\frac{3 h_t^2 m_t^2}{8\pi^2}|\mu|\Big(|A_t|B'_t\,g_t \sin(\phlam' +
  \phi_{A_t}) -
  \frac{|A_t|\sin(\phlam'+\phi_{A_t})}{m_{\tilde{t}_2^2}-m_{\tilde{t}_1^2}}
  \Big) \nn\\ 
&&+\frac{3 h_b^2 m_b^2}{8\pi^2}|\mu|\Big(|A_b|B'_b\,g_b\sin(\phlam' +
  \phi_{A_b}) -
  \frac{|A_b|\sin(\phlam'+\phi_{A_b})}{m_{\tilde{b}_2^2}-m_{\tilde{b}_1^2}}
  \Big) \,, \nn\\ 
{\Delta \cal M}_{SP,13}^2 & = & 
\frac{3 h_b^2 m_b^2|\lambda|v_u(Q)}{8\sqrt{2}\pi^2}\Big(|A_b|B'_b\,g_b\sin(\phlam' +
  \phi_{A_b}) -
  \frac{|A_b|\sin(\phlam'+\phi_{A_b})}{m_{\tilde{b}_2^2}-m_{\tilde{b}_1^2}}
  \Big) \,,\nn\\
{\Delta \cal M}_{SP,23}^2 & = & 
\frac{3 h_t^2 m_t^2|\lambda| v_d(Q)}{8\sqrt{2}\pi^2}\Big(|A_t|B'_t\,g_t \sin(\phlam' +
  \phi_{A_t}) -
  \frac{|A_t|\sin(\phlam'+\phi_{A_t})}{m_{\tilde{t}_2^2}-m_{\tilde{t}_1^2}}
  \Big)\,.
\label{c.18e}
\end{eqnarray}

There are additional $D$-term contributions which are quite involved
but do not give large logarithms since the squarks are assumed to have
masses close to the ultraviolet cutoff $M^2_{\rm SUSY}$. These corrections are given for $\mathcal{M}_S^2$ as
\begin{eqnarray}
{\Delta \cal M}_{S,11}^2 & = &
2|\mu|\Big(|A_t|C_t\cot\beta \cos(\phlam'+\phi_{A_t}) - |A_b|C_b\tan\beta
\cos(\phlam'+\phi_{A_b})\Big) \nn \\
&& + 2|A_b|^2C_b + 2D_b  - 2|\mu|^2C_t\cot^2\beta \,, \nn \\
{\Delta \cal M}_{S,22}^2 & = & 
2|\mu|\Big(|A_b|C_b\tan\beta \cos(\phlam'+\phi_{A_b}) - |A_t|C_t\cot\beta
\cos(\phlam'+\phi_{A_t}) \Big) \nn \\
&& + 2|A_t|^2C_t + 2D_t  - 2|\mu|^2C_b\tan^2\beta \,, \nn \\
{\Delta \cal M}_{S,12}^2 & = & \cot\beta\Big(
(|\mu|^2-|A_t|^2)C_t-D_t\Big) +
\tan\beta\Big((|\mu|^2-|A_b|^2)C_b-D_b\Big) \nn \\
&&- |\mu|\Big(|A_t|C_t(1-\cot^2\beta)\cos(\phlam'+\phi_{A_t}) 
+ |A_b|C_b(1-\tan^2\beta)\cos(\phlam'+\phi_{A_b}) \Big)\,, \nn 
\end{eqnarray}
\begin{eqnarray} 
{\Delta \cal M}_{S,13}^2 & = & \frac{|\lambda|}{\sqrt{2}}\Big(
|A_t|C_tv_d(Q)\cot\beta\cos(\phlam'+\phi_{A_t}) -
|A_b|C_bv_u(Q)\cos(\phlam'+\phi_{A_b})\Big) \nn \\
&&+ \frac{|\lambda|^2}{2}s(Q)\Big(v_u(Q)C_b\tan\beta - v_d(Q)C_t \cot\beta\Big) \,, \nn \\
{\Delta \cal M}_{S,23}^2 & = & \frac{|\lambda|}{\sqrt{2}}\Big(
|A_b|C_bv_u(Q)\tan\beta\cos(\phlam'+\phi_{A_b}) -
|A_t|C_tv_d(Q)\cos(\phlam'+\phi_{A_t})\Big) \nn \\
&&+ \frac{|\lambda|^2}{2}s(Q)\Big(v_d(Q)C_t\cot\beta - v_u(Q)C_b \tan\beta\Big) \,,
\end{eqnarray}
where again the quantities $C_f$ and $D_f$ are defined in
Appendix\,\ref{sec:app-B}, and for $\mathcal{M}_{SP}^2$ as
\begin{eqnarray}
{\Delta \cal M}_{SP,11}^2 & = & 2|\mu|\Big(|A_t|C_t\cot\beta \sin(\phlam'+\phi_{A_t}) - |A_b|C_b\tan\beta \sin(\phlam'+\phi_{A_b}) \Big) \,,\nn \\
{\Delta \cal M}_{SP,22}^2 & = & 2|\mu|\Big (|A_b|C_b\tan\beta
\sin(\phlam'+\phi_{A_b}) - |A_t|C_t\cot\beta \sin(\phlam'+\phi_{A_t}) \Big) \,,\nn \\
{\Delta \cal M}_{SP,12}^2 & = & - |\mu|\Big(|A_t|C_t(1-\cot^2\beta)\sin(\phlam'+\phi_{A_t}) 
+ |A_b|C_b(1-\tan^2\beta)\sin(\phlam'+\phi_{A_b}) \Big)\,,\nn \\
{\Delta \cal M}_{SP,13}^2 & = & \frac{|\lambda|}{\sqrt{2}}\Big(
|A_t|C_tv_d\cot\beta\sin(\phlam'+\phi_{A_t}) -
|A_b|C_bv_u\sin(\phlam'+\phi_{A_b})\Big) \,,\nn \\
{\Delta \cal M}_{SP,23}^2 & = & \frac{|\lambda|}{\sqrt{2}}\Big(
|A_b|C_bv_u\tan\beta\sin(\phlam'+\phi_{A_b}) -
|A_t|C_tv_d\sin(\phlam'+\phi_{A_t})\Big) \,.
\end{eqnarray} 

Finally, neglecting all
terms without two powers of large logarithms, the dominant two-loop
squark contributions to the effective potential can be obtained by integrating
the relevant RG equations. These contributions are given as 
\begin{eqnarray}
{\Delta \cal M}_{S,11}^2 & = & \frac{3 h_b^4 v_d^2(Q)}{256\pi^4} \Big(
\ln^2\left(\frac{M_{\rm SUSY}^2}{m_{t}^2}\right)
(16\,g_3^2-\frac{2}{3}\,g_1^2+3\sin^2\beta\, h_t^2 -3\cos^2\beta\, h_b^2)\nn \\
&&+\Big[\ln^2\left(\frac{M_A^2}{m_{t}^2}\right)-
\ln^2\left(\frac{M_{\rm SUSY}^2}{m_{t}^2}\right)\Big]
(3\sin^2\beta\, h_b^2 +(3\sin^2\beta+1)\,h_t^2)\Big)\;,\nn \\
{\Delta \cal M}_{S,22}^2 & = & \frac{3 h_t^4 v_u^2(Q)}{256\pi^4} \Big(
\ln^2\left(\frac{M_{\rm SUSY}^2}{m_{t}^2}\right)
(16\,g_3^2+\frac{4}{3}\,g_1^2-3\sin^2\beta\, h_t^2 +3\cos^2\beta\, h_b^2)\nn \\
&&+\Big[\ln^2\left(\frac{M_A^2}{m_{t}^2}\right)-
\ln^2\left(\frac{M_{\rm SUSY}^2}{m_{t}^2}\right)\Big]
(3\cos^2\beta\, h_t^2 +(3\cos^2\beta+1)\,h_b^2)\Big)\,.
\label{c.19e}
\end{eqnarray} 

\subsubsection{Chargino/neutralino, gauge boson and dominant stau contributions}

Again, some of the radiative corrections due to
the chargino/neutralino loops can be described by an additional 
shift in $A_\lambda$ on top of the corrections in Eq.\,(\ref{c.17e}),
\begin{equation}
\label{c.23e}
A'_\lambda \to A''_\lambda = A'_\lambda + \frac{1}{16\pi^2} (g_1^2 M_1 +3\,g_2^2
M_2) L_{M_2\mu}\,,
\end{equation}
where $M_1$ and $M_2$ are the soft gaugino masses, which are taken here to
be real. The logarithm $L_{M_2\mu}$ is defined, along with $L_{\mu\nu}$, $L_{\mu}$ and
$L_{\nu}$ used in the following, in Appendix\,\ref{sec:app-B}. 
For the CP-odd block in the Higgs mass matrix, all the radiative
corrections due to chargino/neutralino loops, $\sim g^4$, are included
by the above shift. The remaining contributions to $\mathcal{M}_S^2$ are given as
\begin{eqnarray}
{\Delta \cal M}_{S,11}^2 &=& \frac{1}{16\pi^2} \Big[2g^2m_Z^2
\cos^2\beta (-10+16\sin^2\theta_W-8\sin^4\theta_W) L_{M_2\mu}\nn \\
&&-4\left(|\mu|^2\mathcal{R}\tan\beta
+\frac{\lambda^4m_Z^2\cos^2\beta}{g^2}\right) L_{\mu\nu}
\Big]\,,\nn \\
{\Delta \cal M}_{S,22}^2 &=& \frac{1}{16\pi^2} \Big[2g^2m_Z^2
\sin^2\beta (-10+16\sin^2\theta_W-8\sin^4\theta_W) L_{M_2\mu}\nn \\
&&-4\left(|\mu|^2\mathcal{R}\cot\beta
+\frac{\lambda^4m_Z^2\sin^2\beta}{g^2}\right) L_{\mu\nu} \Big]\,,\nn\\
{\Delta \cal M}_{S,33}^2 &=& \frac{1}{16\pi^2} \Big(
 -32|\kappa|^2\nu^2 L_{\nu} -8 |\lambda|^2|\mu|^2L_{\mu}
\Big)\,,\nn 
\end{eqnarray}
\begin{eqnarray}
{\Delta \cal M}_{S,12}^2 &=& \frac{1}{16\pi^2} \Big[
4\left(|\mu|^2\mathcal{R} - \frac{\lambda^4m_Z^2\sin\beta
\cos\beta}{g^2}\right) L_{\mu\nu} \nn \\
&&-4g^2m_Z^2\sin\beta \cos\beta L_{M_2\mu} \Big]\,, \nn \\
{\Delta \cal M}_{S,13}^2 &=& \frac{1}{16\pi^2}
\frac{m_Z s(Q) }{\sqrt{g_1^2+g_2^2}} \Big[2 |\lambda|^2 g^2
\cos\beta(-6+4\sin^2\theta_W) L_{M_2\mu} \nn \\
&&+2\sqrt{2}|\lambda|^2\Big(2\mathcal{R}\sin\beta - (|\lambda|^2+4|\kappa|^2)\cos\beta\Big)
L_{\mu\nu} \Big]\,, \nn \\
{\Delta \cal M}_{S,23}^2 &=& \frac{1}{16\pi^2}
\frac{m_Z s(Q) }{\sqrt{g_1^2+g_2^2}} \Big[2 |\lambda|^2 g^2
\sin\beta(-6+4\sin^2\theta_W) L_{M_2\mu} \nn \\
&&+2\sqrt{2}|\lambda|^2\Big(2\mathcal{R}\cos\beta - (|\lambda|^2+4|\kappa|^2) \sin\beta\Big)
L_{\mu\nu} \Big]\,,
\label{c.24e}
\end{eqnarray}
where $|\nu| \equiv |\kappa|s(Q)/\sqrt{2}$ and $m_Z$ is the mass of
the $Z$ boson. The corresponding corrections to $\mathcal{M}_{SP}^2$
are given as
\begin{eqnarray}
{\Delta \cal M}_{S,11}^2 &=&-\frac{1}{4\pi^2} |\mu|^2\mathcal{I}\tan\beta L_{\mu\nu}\,, \nn\\
{\Delta \cal M}_{S,22}^2 &=& -\frac{1}{4\pi^2} |\mu|^2\mathcal{I}\cot\beta L_{\mu\nu}\,, \nn\\
{\Delta \cal M}_{S,12}^2 &=& \frac{1}{4\pi^2} |\mu|^2\mathcal{I} L_{\mu\nu}\,, \nn\\
{\Delta \cal M}_{S,13}^2 &=&
\frac{1}{2\sqrt{2}\pi^2}\frac{m_Z s(Q)}{\sqrt{g_1^2+g_2^2}}|\lambda|^2\mathcal{I}\sin\beta
L_{\mu\nu}\,, \nn \\
{\Delta \cal M}_{P,23}^2 &=& 
\frac{1}{2\sqrt{2}\pi^2}\frac{m_Z s(Q)}{\sqrt{g_1^2+g_2^2}}|\lambda|^2\mathcal{I}\cos\beta
L_{\mu\nu}\,.
\end{eqnarray} 

The contributions from gauge bosons can be conveniently written as 
\begin{eqnarray}
{\Delta \cal M}_{S,11}^2 &=& \Delta_\text{Gauge} \cos^2\beta\;, \nn \\
{\Delta \cal M}_{S,22}^2 &=& \Delta_\text{Gauge} \sin^2\beta\;, \nn \\
{\Delta \cal M}_{S,12}^2 &=& \Delta_\text{Gauge} \sin\beta\cos\beta\,,
\label{c.28e}
\end{eqnarray}
in terms of the auxiliary quantity
\begin{eqnarray}
\label{c.27e}
\Delta_\text{Gauge}=\frac{1}{16\pi^2}g^2m_Z^2 
(-9+12\sin^2\theta_W-6\sin^4\theta_W) \ln\left(\frac{M_{\rm SUSY}^2}{m_Z^2}\right)\,.
\end{eqnarray}

Finally, staus can be considerably lighter than
the third generation squarks and hence can give comparatively larger
$D$-term contributions which are written as
\begin{eqnarray}
{\Delta \cal M}_{S,11}^2 &=& \Delta_{\tilde{\tau}} \cos^2\beta\,, \nn \\
{\Delta \cal M}_{S,22}^2 &=& \Delta_{\tilde{\tau}} \sin^2\beta\,, \nn \\
{\Delta \cal M}_{S,12}^2 &=& -\Delta_{\tilde{\tau}} \sin\beta\cos\beta\,,
\label{c.21e}
\end{eqnarray} 
where, assuming a common stau mass, $m_{\tilde{\tau}}$,
\begin{equation}
\label{c.20e}
\Delta_{\tilde{\tau}} = -\frac{1}{16\pi^2}g^2m_Z^2  (9\sin^4\theta_W
+3\cos^4\theta_W) \ln\left(\frac{M_{\rm
SUSY}^2}{m_{\tilde{\tau}}^2}\right) \,,
\end{equation}
with $\theta_W$ being the weak mixing angle. 

\subsubsection{Wave function renomalisation}

As mentioned earlier, the elements of the
loop-corrected Higgs mass matrix obtained so far contain VeVs
$v_u(Q)$, $v_d(Q)$ and $s(Q)$ defined at the scale $Q^2=M^2_{\rm SUSY}$. 
These VeVs are related to the VeVs of the properly normalised Higgs
fields (i.e., after the addition of quantum effects with $Q^2 < M^2_{\rm SUSY}$) as
\begin{eqnarray}
\label{c.12e}
v_u(Q) = \frac{v_u}{\sqrt{Z_{H_u}}}\,,\qquad v_d(Q) = \frac{v_d}{\sqrt{Z_{H_d}}}\,,
\qquad s(Q) = \frac{s}{\sqrt{Z_{S}}}\,,
\end{eqnarray}
where $Z_i$, with $i = H_u,\,H_d,\,S$, are the wave function renormalisation
constants. These constants multiply the kinetic terms in the effective
action and their explicit forms are given in Appendix\,\ref{sec:app-B}. 
The elements of the Higgs mass matrix, therefore, have to be
rescaled by appropriate powers of these renormalization constants as  
\begin{eqnarray}
\label{c.29e}
{\cal M'}_{H,ij}^2 = {\cal M}_{H,ij}^2/\sqrt{Z_i\,Z_j}\,.
\end{eqnarray}
This rescaling then takes care of further contributions of the order $g^2\,
h_{t,b}^2$ to the Higgs mass matrix. 

\subsection{\label{sec:hmasses} Physical Higgs boson masses}
To obtain the physical mass eigenstates the $6\times 6$ Higgs
mass matrix ${\cal M'}_{H}^2$ can be diagonalised using the orthogonal matrix $O$ as 
\begin{eqnarray}
\label{eq:rot66}
(H_1,\,H_2,\,H_3,\,H_4,\,H_5,\,H_6)_a^T = O_{ai}\,
(H_{dR},\,H_{uR},\,S_R,\,H_{dI},\, H_{uI},\,S_I)^T_i \,.
\end{eqnarray} 
However, one of the resulting states corresponds to a massless Nambu-Goldstone (NG)
mode, $G$. In order to isolate this NG mode, a $\beta$ rotation of $\mathcal{M}_P^2$ is carried out, before the above diagonalisation, as
\begin{equation}
\left(\begin{array}{c}H_{dI}\\H_{uI}\\S_I\end{array}\right)
 =\left(\begin{array}{ccc}
 \cos\beta&\sin\beta&0\\
 -\sin\beta&\cos\beta&0\\
 0&0&1
 \end{array}\right)
 \left(\begin{array}{c}G\\H_I\\S_I\end{array}\right)\,.
\end{equation}
In the new basis, ${\bf h}^T\equiv\left(
H_{dR},\,H_{uR},\,S_R,\,H_I,\,S_I\right)$, after dropping the
NG mode, the tree level pseudoscalar block in
Eq.\,(\ref{eq:mhiggs}) gets replaced by 
\begin{eqnarray}
{\cal M}^{2}_{P_\beta}=
\left(
        \begin{array}{cc}
 (R_\lambda+\mathcal{R}s/2)\frac{v^2s}{v_dv_u} 
& (R_\lambda -\mathcal{R}s)v \\
 (R_\lambda -\mathcal{R}s)v 
 & R_\lambda\frac{v_dv_u}{s}+2\mathcal{R}v_dv_u-3R_\kappa s 
                \end{array}
\right)\,,
 \end{eqnarray}
where $v=\sqrt{v_u^2 + v_d^2}$, and the off-diagonal CP-mixing block
gets replaced by
\begin{eqnarray}
{\cal M}^{2}_{SP_\beta}=
\left(
        \begin{array}{cc}
0 & -\frac{3}{2}\mathcal{I}sv_u \\
0 & -\frac{3}{2}\mathcal{I}sv_d \\
\frac{1}{2}\mathcal{I}sv & -2\mathcal{I}v_uv_d 
                \end{array}
\right)\,.
\end{eqnarray}
The radiatively corrected Higgs mass matrix in the new basis
can be obtained by similarly $\beta$-rotating the mass matrix given in
Eq.\,(\ref{c.29e}) as   
\begin{eqnarray}
\label{eq:mhiggs-rot}
\mathcal{M'}^{2}_h = ({\cal M'}_{H}^2)_\beta\,.
\end{eqnarray}
The effective potential masses of the neutral Higgs bosons are then
obtained by diagonalising the above $5\times 5$ mass matrix as
$O'^T{\mathcal{M'}}_h^2O'={\rm diag}
(m^2_{h_1}\;m^2_{h_2}\;m^2_{h_3}\;m^2_{h_4}\;m^2_{h_5})$, such that 
\begin{equation}
  \label{eq:mass-order}
m^2_{h_1}\leq m^2_{h_2}\leq m^2_{h_3} \leq m^2_{h_4} \leq m^2_{h_5}\,.
\end{equation}

For Higgs boson pole masses the approximate expression obtained in
\cite{Ellwanger:2009dp} can be extrapolated to the cNMSSM as
\begin{eqnarray}
\hspace*{-1.0cm}m^{\text{pole}\ 2}_{h_i} &=& m_{h_i}^2 
-\frac{3h_t^2}{16\pi^2}
\Big[(m_{h_i}^2-4m_{t}^2) O_{i2}^2 + m_{h_i}^2O_{i5}^2\Big] B(m_{h_i}^2,m_{t}^2)\nn \\
&-&\frac{3h_b^2}{16\pi^2}\Big[
m_{h_i}^2 (O^2_{i1}+O^2_{i4})\ln\left(\frac{m_{t}^2}{m_{b}^2}\right)
+ \Big((m_{h_i}^2-4m_{b}^2) O_{i1}^2 + m_{h_i}^2O_{i4}^2\Big) B(m_{h_i}^2,m_{b}^2)\Big]\,,
\label{c.36e}
\end{eqnarray}
where the function $B(M^2,m^2)$ is defined as
\begin{eqnarray}
B(M^2,m^2) =\left\{\begin{array}{cl}
           2-\sqrt{1-\frac{4m^2}{M^2}}
\ln\left(\frac{1+\sqrt{1-\frac{4m^2}{M^2}}}
{1-\sqrt{1-\frac{4m^2}{M^2}}}\right) \,:   & \qquad M^2 > 4m^2\,, \\
  2-2\sqrt{\frac{4m^2}{M^2}-1}\arctan
\left(\sqrt{\frac{M^2}{4m^2-M^2}}\right)\,: & \qquad M^2 < 4m^2\,.
\end{array}\right.
\label{c.37e}
\end{eqnarray}

In the case of the charged Higgs states, a $\beta$ rotation is also carried
out for isolating the NG modes. The
corrections to the charged Higgs boson mass, of the order $h_{t,b}^4$ and those
induced by chargino/neutralino, gauge boson and slepton loops give rise to some
additional terms on top of the shifts of $A_\lambda$ described earlier. After including
these corrections and $\beta$-rotating, the mass of the physical
charged Higgs boson is given  as
\begin{eqnarray}
{\cal M'}_\pm^2 &=& \Big[(R_\lambda+\mathcal{R}s/2)s + v_u(Q) v_d(Q) \left(\frac{g_2^2}{2} - |\lambda|^2\right)\Big]
\left(\frac{v_u(Q)}{Z_{H_d}v_d(Q)} + \frac{v_d(Q)}{Z_{H_u}v_u(Q)}\right)\nn \\
&&+\frac{v_u^2(Q)+ v_d^2(Q)}{16\pi^2} 
\Big[ 6 h_t^2 h_b^2 \ln\left(\frac{M_{\rm SUSY}^2}{m_t^2}\right)
-\frac{3}{4} g_2^4 \ln\left(\frac{M_{\rm SUSY}^2}{m_{\tilde{l}}^2}\right)\nn \\
&& + \frac{7g_1^2g_2^2 - g_2^4}{4}
\ln\left(\frac{M_{\rm SUSY}^2}{m_Z^2}\right) + 2(g_1^2g_2^2 - g_2^4)L_{M_2\mu}
\Big]\,,
\label{c.31e}
\end{eqnarray}
where the rescaling by the wave function normalisation constants has
been taken care of. The pole mass of the charged Higgs boson is then obtained as
\begin{eqnarray}\label{c.39e}
m_{h^\pm}^{\text{pole}\ 2} &=& {\cal M'}_\pm^2 
+\frac{3}{16\pi^2}\Bigg\{
(h_t^2\cos^2\beta+h_b^2\sin^2\beta)\Bigg({\cal M'}_\pm^2
\left[\left(1-\frac{m_{t}^2}{{\cal M'}_\pm^2}\right)
\ln\left|\frac{{\cal M'}_\pm^2-m_{t}^2}{m_{t}^2}\right|-2\right]\nn
\\ &&
+(m_{t}^2+m_{b}^2)
\left[\left(1-\frac{m_{t}^2}{{\cal M'}_\pm^2}\right)
\ln\left|\frac{m_{t}^2}{{\cal M'}_\pm^2-m_{t}^2}\right|+1\right]
\Bigg) \nn \\
&& +\frac{4m^2_{t} m^2_{b}}{v^2}
\left[\left(1-\frac{m_{t}^2}{{\cal M'}_\pm^2}\right)
\ln\left|\frac{m_{t}^2}{{\cal M'}_\pm^2-m_{t}^2}\right|+1\right]
\Bigg\}\,.
\end{eqnarray}

\section{\label{sec:couplings} Trilinear Higgs boson
  self-interactions}

The complete NMSSM Lagrangian contains the interaction terms of the
 Higgs bosons with the fermions, scalars and vector bosons as well as
 with each other, from which the corresponding couplings can be obtained.   
In table\,\ref{tab:hcplgs} we
summarize various Higgs boson couplings which will be used 
in the expressions for neutral Higgs
boson decay widths in the next section. The analytical formulae for these
couplings in the cNMSSM, with the exception of the Higgs boson self-couplings,
can be found in \cite{Cheung:2010ba,*Cheung:2011wn}. The
couplings between three neutral Higgs bosons, 
obtained from the potential in Eq.\,(\ref{eq:Higgspot}), are given as
\begin{eqnarray}
g_{h_a h_b h_c} & = &
\frac{g^2}{4} \Big( v_u (\Pi_{abc}^{111} -
\Pi_{abc}^{122} + \Pi_{abc}^{144} - \Pi_{abc}^{155}) + v_d
(\Pi_{abc}^{222} - \Pi_{abc}^{211} + \Pi_{abc}^{255} -
\Pi_{abc}^{244}) \Big) \nonumber \\
& & +\frac{\lambda^2}{2} 
\Big( v_u (\Pi_{abc}^{122}+\Pi_{abc}^{133}
+ \Pi_{abc}^{155} + \Pi_{abc}^{166}) 
 + v_d (\Pi_{abc}^{211}+\Pi_{abc}^{233}+\Pi_{abc}^{244} +
\Pi_{abc}^{266}) \nonumber \\ 
&&+ s(\Pi_{abc}^{311}+\Pi_{abc}^{322}+\Pi_{abc}^{344}+\Pi_{abc}^{355})
\Big)  + \kappa^2 s \Big(\Pi_{abc}^{333} + \Pi_{abc}^{366}\Big)\nonumber \\
&& - R_\lambda \Big(\Pi_{abc}^{123} - \Pi_{abc}^{453} 
- \Pi_{abc}^{426} - \Pi_{abc}^{156}\Big) 
+ R_{\kappa} \Big(\Pi_{abc}^{333} - 3\Pi_{abc}^{366}\Big) \nonumber \\
&&- \frac{\mathcal{R}}{2}\Big( v_u (\Pi_{abc}^{233} -
\Pi_{abc}^{266} + 2 \Pi_{abc}^{536}) 
+ v_d (\Pi_{abc}^{133} -
\Pi_{abc}^{166} + \Pi_{abc}^{436}) \nonumber \\ 
&& + 2 s (\Pi_{abc}^{123} - \Pi_{abc}^{345} + \Pi_{abc}^{156} +
\Pi_{abc}^{426}) \Big) \nonumber \\ 
&& - \frac{\mathcal{I}}{2}\Big( v_u (\Pi_{abc}^{566} -
\Pi_{abc}^{533} + 2 \Pi_{abc}^{236}) + v_d (\Pi_{abc}^{466} -
\Pi_{abc}^{433} + \Pi_{abc}^{136}) \nonumber \\ 
&& + s (3\Pi_{abc}^{126} - 3\Pi_{abc}^{456} - \Pi_{abc}^{135} -
\Pi_{abc}^{423}) + 3\frac{v_d v_u}{s}(\Pi_{abc}^{666}-\Pi_{abc}^{336}) \Big) \,, 
\end{eqnarray}
where
\begin{eqnarray}
\hspace*{-0.4cm} \Pi_{abc}^{ijk} & = &
O_{ai} O_{bj} O_{ck} + O_{ai} O_{cj} O_{bk} + O_{bi} O_{aj} O_{ck}
+ O_{bi} O_{cj} O_{ak} + O_{ci} O_{aj} O_{bk} + O_{ci} O_{bj} O_{ak}\,,
\end{eqnarray}
with $O_{xy}$ being the elements of the Higgs mixing matrix defined
in Eq.\,(\ref{eq:rot66}). 
The couplings of the neutral Higgs bosons to a pair of charged Higgs
bosons are similarly given as
\begin{eqnarray}
g_{h_a h^+ h^-} & = &
\frac{g_1^2}{8} \Big( v_u (\Pi_{abc}^{111} -
\Pi_{abc}^{122}) + v_d
(\Pi_{abc}^{222} - \Pi_{abc}^{211}) \Big) \nonumber \\
& & + \frac{g_2^2}{8} \Big( v_u (\Pi_{abc}^{111} +
\Pi_{abc}^{122} + 2\Pi_{abc}^{212}) + v_d
(\Pi_{abc}^{222} + \Pi_{abc}^{211} + 2\Pi_{abc}^{112}) \Big) \nonumber
\\ 
& & +\frac{\lambda^2}{2} 
\Big( s (\Pi_{abc}^{311}+\Pi_{abc}^{322})
 - v_u \Pi_{abc}^{212} - v_d \Pi_{abc}^{112} \Big) \nonumber \\ 
&& + \mathcal{R}s \Pi_{abc}^{312} 
+ R_\lambda \Pi_{abc}^{312} +\frac{3}{2} \mathcal{I}s\Pi_{abc}^{612}\,,
\end{eqnarray}
where
\begin{equation}
\Pi_{abc}^{ijk}  = 
2O_{ai} C_jC_k \; {\rm with}\;C_1=\cos\beta\,,\,C_2=\sin\beta\,.
\end{equation}

\begin{table}[t]
\begin{center}
\begin{tabular}{|c|cc|}
\hline
{\it fermion pair} &  $g_{h_a\bar{f}f}^S$  & $g_{h_a\bar{f}f}^P$    \\
\hline
$d \bar{d}/l^+ l^-\,$   & $O_{a1}/\cos\beta$  & $-O_{a4}/\cos\beta $ \\
$u \bar{u}\,$ & $O_{a2}/\sin\beta$  & $-O_{a5}/\sin\beta $ \\
$ \tilde{\chi}_j^0 \tilde{\chi}_k^0\,$
 & $g_{h_a\tilde{\chi}^0_j\tilde{\chi}^0_k}^{S}$
 & $g_{h_a\tilde{\chi}^0_j\tilde{\chi}^0_k}^{P}$ \\
$ \tilde{\chi}_j^- \tilde{\chi}_k^+\,$
 & $g_{h_a\tilde{\chi}^+_j\tilde{\chi}^-_k}^{S}$
 & $g_{h_a\tilde{\chi}^+_j\tilde{\chi}^-_k}^{P}$ \\
\hline
\hline
{\it sfermions} & \multicolumn{2}{|c|}{$g_{h_a \tilde{f}_b\widetilde{f}_c^*}$}  \\
{\it gauge bosons} & \multicolumn{2}{|c|}{$g_{h_a V V}$}\\
{\it Higgs\,+\,Z boson} & \multicolumn{2}{|c|}{$g_{h_a h_bZ}$} \\
\hline
{\it neutral Higgs bosons} & \multicolumn{2}{|c|}{$g_{h_a h_b h_c}$} \\
{\it charged Higgs bosons} & \multicolumn{2}{|c|}{$g_{h_a h^+h^-}$} \\
\hline
\end{tabular}
\caption{\label{tab:hcplgs}
The couplings of the NMSSM Higgs boson $h_a$ to particles and
sparticles at the tree level. $g_{h_a\bar{f}f}^S$ and $g_{h_a\bar{f}f}^P$ refer to vector and
axial vector couplings of the fermions, respectively. $O_{ai}$ are
the elements of the Higgs mixing matrix defined in Sect.\,\ref{sec:hmasses}.
}
\end{center}
\end{table}

\section{\label{sec:decays} Neutral Higgs boson decays}

In this section, we present the analytical expressions for the 
decay widths of the cNMSSM Higgs bosons into pairs
of fermions, massive gauge bosons, sfermions, photons, gluons
and lighter Higgs bosons as well as into a lighter Higgs and 
massive gauge boson pair. These expressions have mostly been
adopted from\,\cite{Lee:2003nta} and follow the notation therein. For
the decay modes involving an off-mass-shell gauge boson, three-body
decays are described following\,\cite{Spira:1997dg}. \\

\noindent $\bullet$ \underline{$h \rightarrow f f^\prime$} \\

\noindent The decay width of a Higgs boson into two fermions is given as
\begin{eqnarray}
\hspace{-0.5 cm}
\Gamma(h_a\rightarrow f f^\prime)&=& N_C
\frac{G_F M_{h_a} \lambda^{1/2}(1,\kappa_{af},\kappa_{af'})}{4\sqrt{2}\pi}
\bar{m}^2_q(m_{h_a}) \Gamma_M K^f_a\nn\\
&&\times
\left[(1\!-\!\kappa_{af}\!-\!\kappa_{af'})(|g_{h_aff'}^S|^2+|g_{h_aff'}^P|^2)\right.\nn\\
&&\left. -2\sqrt{\kappa_{af}\kappa_{af'}} (|g_{h_aff'}^S|^2-|g_{h_aff'}^P|^2)\right]\,,
\end{eqnarray}
where $G_F/\sqrt{2}=g_2^2/8m_W^2$, with $m_W$ being the $W$ boson mass,
 $\kappa_{af^{(')}} \equiv m_{f^{(')}}^2/m_{h_a}^2$,
$\lambda(1,x,y) \equiv (1-x-y)^2-4xy$ and the couplings $g_{h_aff'}^S$
and $g_{h_aff'}^P$ have been defined in table\,\ref{tab:hcplgs}
(where $f'=\bar{f}$ in the case of quarks and leptons). The colour factor
$N_C$ is equal to 3 for quarks and to 1 for leptons, charginos and neutralinos.
$\Gamma_M= \left(\frac{4}{1+\delta_{bc}}\right)$ for Majorana
fermions such as (s)neutrinos, neutralinos and charginos, with
$\delta_{bc}=1$ when they are identical, while for Dirac fermions $\Gamma_M=1$.
For $h_a \rightarrow q\bar{q}$ the leading-order QCD corrections are
taken into account with the enhancement factor
$K^q_a = 1+5.67\,\frac{\alpha_s(m_{h_a}^2)}{\pi}$. For
leptons, neutralinos and charginos, $K^f_a = 1$. \\

\noindent $\bullet$ \underline{$h \rightarrow VV$} \\

\noindent The decay width into two massive gauge bosons is given as
\begin{eqnarray}
\Gamma(h_a\rightarrow VV) &=& \delta_V\frac{G_F\,g_{_{h_aVV}}^2 m_{h_a}^3}{16\sqrt{2}\pi}
\beta_{iV}(1-4\kappa_{aV}+12\kappa_{aV}^2)\,,
\end{eqnarray}
where $\kappa_{aV}=m_V^2/m_{h_a}^2$,
$\beta_{aV}=\sqrt{1-4\kappa_{aV}}$,
$\delta_W=2$ and $\delta_Z=m_W^4/(\cos\theta_W\,m_Z)^4=1$. 
Below the $VV$ threshold, when one of
the gauge bosons is off mass shell, the three-body decay width of a
Higgs boson is given as
\begin{equation}
\Gamma(h_a\rightarrow VV^*)=\delta'_V\frac{3G_F\,g_{_{h_aVV}}^2
  m_{h_a} m_V^4}{16\sqrt{2}\pi} R(\kappa_{aV})
\,,
\end{equation}
where $\delta'_W=2$, $\delta'_Z=7/12 - 10\sin^2\theta_W/9 + 40
\sin^4\theta_W /27$ and
\begin{equation}
R(x) = 3 \frac{1-8x+20x^2}{\sqrt{4x-1}} \arccos \left(\frac{3x-1}{2x^{3/2}}
\right) - \frac{1-x}{2x} (2-13x+47x^2) - \frac{3}{2} (1-6x+4x^2) \log x \,.
\end{equation} \\

\noindent $\bullet$ \underline{$h_a\rightarrow h_bh_c$, $\tilde{f}_b\widetilde{f}_c^*$} \\

\noindent The decay width of a Higgs boson into two scalar particles, including
sfermions and lighter Higgs bosons, is written as
\begin{equation}
\label{eq:htoh-dec}
\Gamma(h_a\rightarrow h_bh_c,\,\tilde{f}_b\widetilde{f}_c^*)\ =\ \frac{N_F |{\cal G}|^2}{16\pi
m_{h_a}}\lambda^{1/2}(1,\kappa_{ab},\kappa_{ac})\,,
\end{equation}
where $(N_F,{\cal G}) =
(1/(1+\delta_{bc}),\,g_{h_ah_bh_c})$, $(N_C,\,
g_{h_a\tilde{f}_b\widetilde{f}_c^*})$ and
$\kappa_{ai}=m_{h_i,\tilde{f}_i}^2/m_{h_a}^2$. \\

\noindent $\bullet$ \underline{$h_a\rightarrow h_b Z$} \\

\noindent The decay width of a Higgs boson into a lighter
Higgs and $Z$ boson pair is given as
\begin{equation}
\label{eq:hhhwidth}
\Gamma(h_a\rightarrow h_b Z) \ = \
g_{h_a h_b Z}^2 \frac{G_F m_Z^4}{8\sqrt{2}\pi m_{h_a}}
\sqrt{\lambda'(m_{h_b}^2,m_Z^2;m_{h_a}^2)} \lambda'(m_{h_b}^2,m_{h_a}^2;m_Z^2) \,,
\end{equation}
where the function $\lambda'(x,y;z) = (1-x/z-y/z)^2
- 4xy/z^2$. Below the threshold for the above process, the three-body decay width is given as
\begin{equation}
\Gamma (h_a\to h_bZ^*)  =  g_{h_a h_b Z}^2 \delta'_Z
\frac{9G_F^2m_Z^4 m_{h_a}}{16\pi^3} G_{h_bZ}\,,
\end{equation}
where the generic functions $G_{ij}$ can be written as
\begin{eqnarray}
G_{ij} & = & \frac{1}{4} \left\{ 2(-1+\kappa_j-\kappa_i)\sqrt{\lambda_{ij}}
\left[ \frac{\pi}{2} + \arctan \left(\frac{\kappa_j (1-\kappa_j+\kappa_i) -
\lambda_{ij}}{(1-\kappa_i) \sqrt{\lambda_{ij}}} \right) \right]
\right. \nonumber \\
& & \left. + (\lambda_{ij}-2\kappa_i) \log \kappa_i + \frac{1}{3} (1-\kappa_i)
\left[ 5(1+\kappa_i) - 4\kappa_j + \frac{2}{\kappa_j} \lambda_{ij} \right]
\right\}\,,
\end{eqnarray}
using the parameters
\begin{equation}
\lambda_{ij} = -1+2\kappa_i+2\kappa_j-(\kappa_i-\kappa_j)^2; \hspace{2cm}
\kappa_i = \frac{m_i^2}{m_{h_a}^2} \, .
\end{equation} \\

\noindent $\bullet$ \underline{$h\rightarrow Z\gamma$} \\

\noindent The decay width of a Higgs boson into a $Z$ boson and photon
pair is given by
\begin{eqnarray}
\Gamma(h_a\rightarrow Z\gamma)=\frac{G_Fm_W^2\alpha^2 m_{h_a}^3}{64\pi^3}
         (1-\kappa_{aZ})^3\left[\,\left|S^{Z\gamma}_a(m_{h_a})\right|^2
              +\left|P^{Z\gamma}_a(m_{h_a})\right|^2\right]\,.
\end{eqnarray}
In the above expression the scalar and
pseudoscalar form factors, retaining only the dominant loop
contributions, which include those from $t$ and $b$ quarks, $W^\pm$
and $H^\pm$, are given by
\begin{eqnarray}
S^{Z\gamma}_a(m_{h_a})&=&\sum_{f=b,t} 
g^{S}_{h_a\bar{f}f}\,F'_{sf}(\tau_{af},\lambda_f) 
+ g_{h_aVV}F'_1(\tau_{aW},\lambda_W) \nonumber \\
&& \nonumber \\
&&
+ \frac{g_{h_ah^+h^-}}{2\sqrt{2}G_Fm_{h^\pm}^2} F'_0(\tau_{ah^\pm},\lambda_{h^\pm})
\,, \nonumber \\
P^{Z\gamma}_a(m_{h_a})&=&\sum_{f=b,t}
g^{P}_{h_a\bar{f}f} \,F'_{pf}(\tau_{af},\lambda_f)
 \,,
\end{eqnarray}
where $\tau_{ax}=4m_x^2/m_{h_a}^2$ and $\lambda_x=4m_x^2/m_Z^2$. The form factors $F'_{sf}$,
$F'_{pf}$, $F'_0$ and $F'_1$ are given as
\begin{eqnarray}
F'_{sf} (\tau,\lambda) & = & 6\frac{Q_f (I_{3f} - 2Q_f\sin^2\theta_W )}
{\cos\theta_W} \left[I_1(\tau,\lambda) - I_2(\tau,\lambda)
\right]\,, \nonumber \\
 F'_{pf} (\tau,\lambda) & = & 12\frac{Q_f (I_{3f} - 2Q_f\sin^2\theta_W )}
{\cos\theta_W}~I_2(\tau,\lambda) \,,\nonumber \\
F'_0 (\tau,\lambda) & = & \frac{\cos 2\theta_W}{\cos\theta_W}
I_1(\tau,\lambda)\,,  \nonumber \\
 F'_1 (\tau,\lambda) & = & \cos \theta_W \left\{ 4(3-\tan^2\theta_W)
I_2(\tau,\lambda) \right. \nonumber \\
& & \left. + \left[ \left(1+\frac{2}{\tau}\right) \tan^2\theta_W
- \left(5+\frac{2}{\tau} \right) \right] I_1(\tau,\lambda) \right\} \,,
\label{eq:hzgaform}
\end{eqnarray}
where $Q_f$ is the electric charge of
the fermion $f$ and $I_{3f}$ is the third component of its
isospin. The functions $I_{1,2}$ are defined as
\begin{eqnarray}
I_1(\tau,\lambda) & = & \frac{\tau\lambda}{2(\tau-\lambda)}
+ \frac{\tau^2\lambda^2}{2(\tau-\lambda)^2} \left[ f(\tau) - f(\lambda) \right]
+ \frac{\tau^2\lambda}{(\tau-\lambda)^2} \left[ g(\tau) - g(\lambda) \right] \,,\\
I_2(\tau,\lambda) & = & - \frac{\tau\lambda}{2(\tau-\lambda)}\left[ f(\tau)
- f(\lambda) \right]\,,
\end{eqnarray}
with
\begin{equation}
g(\tau) = \left\{ \begin{array}{ll}
\displaystyle \sqrt{\tau-1} \arcsin \frac{1}{\sqrt{\tau}} & \tau \ge 1 \\
\displaystyle \frac{\sqrt{1-\tau}}{2} \left[ \log \frac{1+\sqrt{1-\tau}}
{1-\sqrt{1-\tau}} - i\pi \right] & \tau < 1\,.
\end{array} \right.
\label{eq:gtau}
\end{equation} 

\noindent $\bullet$ \underline{$h\rightarrow \gamma\gamma$} \\

\noindent The decay width into two photons is given as
\begin{eqnarray}
\Gamma(h_a\rightarrow \gamma\gamma)=\frac{G_F\alpha^2 m_{h_a}^3}{128\sqrt{2}\pi^3}
         \left[\,\left|S^\gamma_a(m_{h_a})\right|^2
              +\left|P^\gamma_a(m_{h_a})\right|^2\right]\,,
\end{eqnarray}
where $\alpha$ is the fine-structure constant. The scalar and
pseudoscalar form factors, retaining only the loop
contributions from $W^\pm$, $H^\pm$ and the dominant ones
from (s)fermions, are given by
\begin{eqnarray}
S^\gamma_a(m_{h_a})&=&2\sum_{f=b,t,\tilde{\chi}^\pm_1,\tilde{\chi}^\pm_2} N_C\, Q_f^2\,
g^{S}_{h_a\bar{f}f}\,F_{sf}(\tau_{af}) \nonumber \\
&&
+ \sum_{\tilde{f}_j=\tilde{t}_1,\tilde{t}_2,\tilde{b}_1,\tilde{b}_2,
           \tilde{\tau}_1,\tilde{\tau}_2}
N_C\, Q_f^2\frac{g_{H_i\tilde{f}^*_j\tilde{f}_j}}
{2\sqrt{2}G_Fm_{\tilde{f}_j}^2} F_0(\tau_{a\tilde{f}_j})
\nonumber \\
&&+\, g_{h_aVV}F_1(\tau_{aW}) +
\frac{g_{h_ah^+h^-}}{2\sqrt{2}G_Fm_{h^\pm}^2} F_0(\tau_{ah^\pm})
\,, \nonumber \\
P^\gamma_a(m_{h_a})&=&2\sum_{f=b,t,\tilde{\chi}^\pm_1,\tilde{\chi}^\pm_2}
N_C\,Q_f^2\,g^{P}_{h_a\bar{f}f}
\,F_{pf}(\tau_{af})
 \,.
\end{eqnarray}
The form factors $F_{sf}$, $F_{pf}$, $F_0$ and $F_1$ in the above equations are given as
\begin{eqnarray}
F_{sf}(\tau)&=&\tau\,[1+(1-\tau) f(\tau)]\,,~~
F_{pf}(\tau)=\tau\,f(\tau)\,,\\
F_0(\tau)&=&-\tau\,[1-\tau f(\tau)]\,, \hspace{1.1 cm}
F_1(\tau)=-[2+3\tau+3\tau (2-\tau)f(\tau)] \,,\nonumber
\label{formfactor}
\end{eqnarray}
in terms of the scaling function $f(\tau)$ written as
\begin{eqnarray}
f(\tau)=\left\{\begin{array}{cl}
           {\rm arcsin}^2(\frac{1}{\sqrt{\tau}}) \,:   & \qquad \tau\geq 1\,, \\
   -\frac{1}{4}\left[\ln \left(\frac{1+\sqrt{1-\tau}}{1-\sqrt{1-\tau}}\right)
                    -i\pi\right]^2\,: & \qquad \tau < 1\,.
\end{array}\right.
\end{eqnarray} \\

\noindent $\bullet$  \underline{$h \rightarrow gg$} \\

\noindent 
The decay width of a Higgs boson into two gluons is given by
\begin{eqnarray}
\label{eq:hggwidth}
\Gamma(h_a\rightarrow gg)\ =\ \frac{G_F\alpha^2_Sm_{h_a}^3}{16\sqrt{2}\pi^3}
         \left[\,K^g_{S}\, \left|S^g_a(m_{h_a})\right|^2\:
              +\: K^g_{P}\, \left|P^g_a(m_{h_a})\right|^2\right]\,,
\end{eqnarray}
where $\alpha_S$ is the strong coupling constant and the scalar and pseudoscalar form factors,
retaining only the contributions from third generation (s)quarks, are given by
\begin{eqnarray}
S^g_a(m_{h_a})&=&\sum_{f=b,t}
g^{S}_{h_af\bar{f}}\,F_{sf}(\tau_{af})
+ \sum_{\tilde{f}_j=\tilde{t}_1,\tilde{t}_2,\tilde{b}_1,\tilde{b}_2}
\frac{g_{h_a\tilde{f}^*_j\tilde{f}_j}}
{4\sqrt{2}G_Fm_{\tilde{f}_j}^2} F_0(\tau_{a\tilde{f}_j}) \,, \nonumber \\
P^g_a(m_{h_a})&=&\sum_{f=b,t}
g^{P}_{h_af\bar{f}}\,F_{pf}(\tau_{af}) \,,
\end{eqnarray}
with functions $F_{sf}$, $F_{pf}$ and $F_0$, being the same as for the
$\gamma\gamma$ mode above. $K^g_{S,P}$ in Eq.\,(\ref{eq:hggwidth}) are 
QCD loop enhancement factors that include the
leading-order QCD corrections. In the heavy-quark limit, the factors
$K^g_{H,A}$ are given by \cite{Spira:1995rr}
\begin{eqnarray}
  \label{KgHA}
K^g_S &=& 1\ +\ \frac{\alpha_S (M^2_{H_i})}{\pi}\,
\bigg(\,\frac{95}{4} \: -\: \frac{7}{6}\,N_F\,\bigg)\,,\nonumber\\
K^g_P &=& 1\ +\ \frac{\alpha_S (M^2_{H_i})}{\pi}\,
\bigg(\,\frac{97}{4} \: -\: \frac{7}{6}\,N_F\,\bigg)\,,
\end{eqnarray}
where $N_F$ is the number of quark flavours lighter than the $h_a$ boson.\\

\section{\label{sec:results} Novel heavy Higgs boson decays into SM-like 125\,GeV states}

In this section we discuss a cNMSSM scenario which,
if probed at the LHC, could provide an indication of not only the existence
of CP violation in the Higgs sector but also of a non-minimal nature of SUSY.    
For a numerical analysis of this scenario, we use a fortran program
(available on request) in which the Higgs mass matrix 
calculated above has been implemented along with other SUSY mass matrices
(given in Appendix\,\ref{sec:app-A}). This program 
computes the particle mass spectrum for a given set of
the cNMSSM input parameters defined at $M_{\rm SUSY}$. In addition,
all the expressions for decay widths, as given in the previous section, have
been implemented in the program, enabling it 
to also calculate Higgs boson BRs in various decays modes. In the current
version of the program, QCD corrections have been included only in 
the decays into quarks and gluons via $K$-factors, as noted in
eqs.\,(46) and (67) respectively. In the CPC limit, the Higgs boson
masses and BRs have been compared with those given by 
NMSSMTools-v3.2.4\,\cite{NMSSMTools} (with the flag for precision in the calculation
of Higgs boson masses set to the default value of 0). While the mass calculations have been found
to differ by $\sim$ 1\% at the most between the two programs, the differences
in BRs can reach as high as $\sim$ 5\% for some points. This is
mainly because of a more robust treatment of QCD corrections in NMSSMTools, which is
not straightforwardly extendable to the CPV case. 

We also note here that the extension of further corrections 
to the Higgs boson masses, those from Higgs loop contributions and those calculated
in \cite{Degrassi:2009yq} for the real NMSSM (included in NMSSMTools by setting the Higgs
boson mass precision flag to 2), to the cNMSSM in the effective
potential approach is a work in progress. 
However, while such improved precision may slightly alter the
regions of the model parameter space yielding the correct mass of the signal
candidate Higgs boson, the results obtained here for our scenario of interest, 
which is a generic feature of the cNMSSM, should still largely be valid in
those regions.

Our package also tests the output of a given point
 in the cNMSSM parameter space against the constraints from the
 direct searches of the SM (and SUSY) Higgs boson(s) as well as third generation
 squark, stau and light chargino at the large electron positron (LEP) collider. Although no limits from
 $b$-physics, LHC SUSY searches or from relic density measurements 
have so far been implemented in the package, in our current analysis we confine 
ourselves to points from among those which have been found to best comply with 
such constraints (see, e.g.,
\cite{Ellwanger:2011aa,Gunion:2012gc,King:2012is,*Gunion:2012zd,*Benbrik:2012rm,*King:2012tr,*Kowalska:2012gs,*Gherghetta:2012gb,*Barbieri:2013hxa,Ellwanger:2012ke}). 

In the experimental searches the magnitude of the signal is
typically characterized by the `signal strength', $\mu(X) \equiv
\sigma_{\rm obs}(X)/\sigma_{h_{\rm SM}}(X)$, where $h_{\rm SM}$ 
implies a SM Higgs boson with a mass equal to the measured 
one of the observed boson decaying via a given channel $X$.
The theoretical counterpart of this quantity,
sometimes referred to as the {\it reduced cross section}, for a Higgs
boson, $h_i$, produced in the dominant gluon fusion mode is given as 
\begin{eqnarray}
\label{eq:Rxsct1}
\mu_{h_i}(X) = \frac{\sigma(gg\rightarrow h_i)}{\sigma(gg\rightarrow
  h_{\rm SM})}\times \frac{{\rm BR}(h_i\rightarrow X)}{{\rm BR}(h_{\rm
    SM} \rightarrow X)}\,.
\end{eqnarray}
To a good approximation, the
ratio of the production cross sections $\sigma$ of $h_i$ 
and $h_{\rm SM}$ in the above expression can be
substituted by the ratio of their respective decay widths into two gluons. We, therefore, redefine the reduced cross section as
\begin{eqnarray}
\label{eq:Rxsct1}
R_{h_i} (X) \equiv \frac{\Gamma(h_i\rightarrow gg)}{\Gamma(h_{\rm SM}\rightarrow
  gg)}\times \frac{{\rm BR}(h_i \rightarrow X)}{{\rm BR}(h_{\rm SM} \rightarrow X)}\,,
\end{eqnarray}
which is calculated by the program for each of the Higgs bosons of the model. For the
Higgs bosons that are assumed to have escaped detection 
so far, $R_{h_i}(X)$ is tested against the LHC exclusion limit on
$\mu(X)$ wherever it is available for a given decay channel. In case two Higgs bosons of the model are
so close in mass that the event excesses due to each of them cannot be independently
resolved by the experiment, $R_{h_i}(X)$ is simply taken to be the sum of
their individual reduced cross sections.

As noted earlier, the presence of non-zero CPV phases in the Higgs sector of
the NMSSM can result in some unique scenarios which are not possible when CP
is conserved. In particular, a decrease in the mass of a given Higgs
boson with a variation in CPV phases can result in the kinematical opening of new
decay channels. Conversely, a gradual decrease in the Higgs boson mass
can result in the closing of a particular decay channel beyond a
certain value of a given CPV phase, thereby causing
a notable reduction in its total width and a deviation in its
BRs from the CPC case. Indeed such deviations were observed for a
$\sim125$\,GeV SM-like Higgs boson in
\cite{Moretti:2013lya}, owing sometimes to the contribution of the CPV phases to
the gaugino masses also besides the Higgs boson mass itself. Another
crucial possibility arises due to the fact that the Higgs mass
eigenstates do not carry a definite CP assignment for non-zero CPV phases. Hence, couplings
between pseudoscalar and scalar states which are forbidden in the CPC
limit become possible upon the introduction of such phases,
resulting in some `unconventional' Higgs boson decays. 

In the NMSSM, in analogy with the decoupling regime of the MSSM,
when one of the CP-even Higgs bosons is required to have exactly SM-like
couplings and a mass around
125\,GeV the other doublet-like scalar and pseudoscalar Higgs bosons
are typically very heavy, $\gtrsim 500$\,GeV. In this case a correlation
exists between the masses of the light doublet-like and the singlet-like
scalar Higgs bosons such that the latter is either lighter than the
former, in a small portion of the parameter space, or decoupled 
like the other heavy doublet-like Higgs bosons.
On the other hand, the mass of the singlet-like pseudoscalar,
typically $a_1$, approximated at the leading order (for large
$\tan\beta$) by 
\begin{equation} 
m_{a_1}^2 \simeq -\kappa s A_\kappa\,,
\end{equation}
can vary much more freely depending on the size of the parameter
$A_\kappa$, with marginal effect on the masses of the other Higgs bosons. 
It is thus possible for $a_1$ to have a mass
close to twice that of the SM-like Higgs boson. Note also the fact that the
partial decay width of a given Higgs boson, $h_a$, into two
lighter Higgs bosons, given in Eq.\,(\ref{eq:hhhwidth}), is inversely proportional to
$m_{h_a}$. Hence, when CPV phases are
turned on, the decay amplitude of a (now CP-indefinite) $\sim 250$\,GeV Higgs
boson into a pair of SM-like Higgs bosons is non-vanishing. Evidently, lower mass of $a_1$
also implies the availability of more, albeit still rather small,
phase space for its production. 

In the following we will further discuss 
the representative points of three benchmark
cNMSSM parameter space cases wherein
not only a SM-like $\sim 125$\,GeV Higgs boson but also the above
mentioned $\sim 250$\,GeV Higgs boson can be obtained. We will analyse
in detail the impact of variation in the CPV phase
$\phi^\prime_\kappa$ (we fix $\varphi$ to $0^\circ$ so that
 $\phi^\prime_\kappa = \phi_\kappa$)\footnote{Since only the
  difference $\phi^\prime_\lambda-\phi^\prime_\kappa$ enters the Higgs mass matrix at the tree level,
  the variation in Higgs boson properties with varying $\phi^\prime_\kappa$ is
  almost identical to that with varying $\phi^\prime_\lambda$, as was noted in
  \cite{Moretti:2013lya}. However, since $\phi^\prime_\kappa$ is virtually unconstrained by
  the measurements of fermionic EDMs\,\cite{Cheung:2010ba,*Cheung:2011wn,Graf:2012hh}, we only vary this phase in our analysis.
  Also, since \phtri\ does not contribute directly to the Higgs-to-Higgs decay width,
  its relevance to our scenario under consideration is minimal.} on
the properties of the relevant Higgs bosons for these points. We should
indicate here that the chosen points exhibiting our scenario of interest are indeed not
isolated ones and dedicated scans of their neighbourhoods in the
model parameter space
should reveal many more similar points. However, such scans are beyond
the scope of this article since our aim here is to highlight some 
specific characteristics of the parameter regions yielding our representative points, 
rather than to map out their sizes. 
For convenience, we shall refer to the singlet-like pseudoscalar(-like) Higgs boson
generically as $h_p$, to
the $\sim 125$\,GeV SM-like Higgs boson as $h_d$ and to the other
singlet-dominated scalar(-like) boson as $h_s$ henceforth. 

In principle, since the coupling of $h_p$ to a pair of $h_d$ is 
only induced by CPV phases one can expect the corresponding partial decay width
and BR to be minimal. However, as noted
above, the fact that the mass of $h_p$ lies much closer to the
$h_dh_d$ production threshold than
that of the heavy doublet-like Higgs bosons is crucial and provides a 
unique possibility in the context of Higgs boson phenomenology at the LHC. 
Therefore, for quantifying the magnitude of the process where $h_p$ 
produced via gluon fusion decays into one or more $h_d$ which
subsequently decay in the
channel $X$, we compute, following Eq.\,(\ref{eq:Rxsct1}), the auxiliary quantity 
\begin{eqnarray}
\label{eq:Axsct1}
A^{h_p}_{i} (\gamma\gamma) \equiv  \frac{\Gamma(h_p\rightarrow
  gg)}{\Gamma(h_{\rm SM}\rightarrow gg)}\times {\rm BR}(h_p \rightarrow h_d h_i) \times \frac{{\rm BR}(h_d \rightarrow \gamma\gamma)}{{\rm BR}(h_{\rm SM} \rightarrow \gamma\gamma)},
\end{eqnarray}
where $h_{\rm SM}$ refers to a SM Higgs boson with the same mass as
$h_d$. $i=d,s$ in the above equation, since the decay $h_p \rightarrow h_d
h_s$ is also possible when $m_{h_s}< m_{h_p}-m_{h_d}$. The second
of the two Higgs bosons thus produced, whether $h_d$ or $h_s$, is
assumed to have escaped undetected in the recent run of LHC, since no Higgs pair
production has been observed there. It can, however, be
probed mainly in the $b\bar{b}$ decay channel, as discussed in
\cite{Ellwanger:2013ova}, in the next LHC run with $\sqrt{s}=14$\,TeV. We, therefore,
also calculate the corresponding auxiliary rate for this Higgs boson in the $b\bar{b}$
channel. Evidently both $A^{h_p}_{d}(\gamma\gamma)$ and
$A^{h_p}_{s}(\gamma \gamma)$ are by
definition zero in the CPC limit. We stress here that the above
expression gives only a crude estimate of diphoton production rate
via this channel, since the incoming gluons will require a larger
momentum fraction for producing the heavier $h_p$ than for $h_{\rm
  SM}$ and thus their structure functions will differ. However, while
a calculation of the actual total cross section for the process $h_p
\rightarrow h_dh_i \rightarrow X_1X_2$ is needed for
an accurate estimate of its significance at the LHC, the above
expression provides a reasonably good approximation since $h_p$ 
in our scenario of interest is not much
heavier than $h_d$.  Evidently then, such an auxiliary signal rate
cannot be defined for the other, much heavier, Higgs bosons of the model. 

Furthermore, in our analysis below we will compute $R_{h_d} (X)$,
defined in Eq.\,(\ref{eq:Rxsct1}), for $X=\gamma\gamma, ZZ,
\tau^+\tau^- $\footnote{A $\sim 4\sigma$ evidence of a $\sim 125$\,GeV Higgs
boson has now also been established in the $\tau^+\tau^-$
channel\,\cite{ATLAS-CONF-2013-108,CMS-PAS-HIG-13-004}.} for each
benchmark case as a measure of the deviation of $h_d$ from SM-like
properties. $R_{h_d}(X) =1$ thus implies that $h_d$ has an exactly
SM-like signal strength in the channel $X$. As for the SUSY inputs, we
will impose the mSUGRA-inspired unification conditions,
\begin{center}
 $M_0 \equiv M_{Q_3} = M_{U_3} = M_{D_3} = M_{L_3} = M_{E_3} = M_{\rm SUSY}$, \\
 $M_{1/2} \equiv 2M_1 = M_2 =  \frac{1}{3} M_3$,  \\
 $A_0 \equiv A_t = A_b = A_{\tau}$, \\
\end{center}
where $M^2_{\widetilde{Q}_3},\,M^2_{\widetilde{U}_3},\,M^2_{\widetilde{D}_3}$
and $M^2_{\widetilde{L}_3},\,M^2_{\widetilde{E}_3}$ are the soft
SUSY-breaking squared masses
of the third generation squarks and sleptons, respectively. 
Finally, we will fix $\rm{sign}[cos(\phlam + \phi_{A_\lambda})] =
\rm{sign}[cos(\phkap + \phi_{A_\kappa})] = +1$. 

\subsection{$h_1 = h_d$}

\begin{figure}[t]
\centering
\subfloat[]{%
\includegraphics*[angle=-90,scale=0.285]{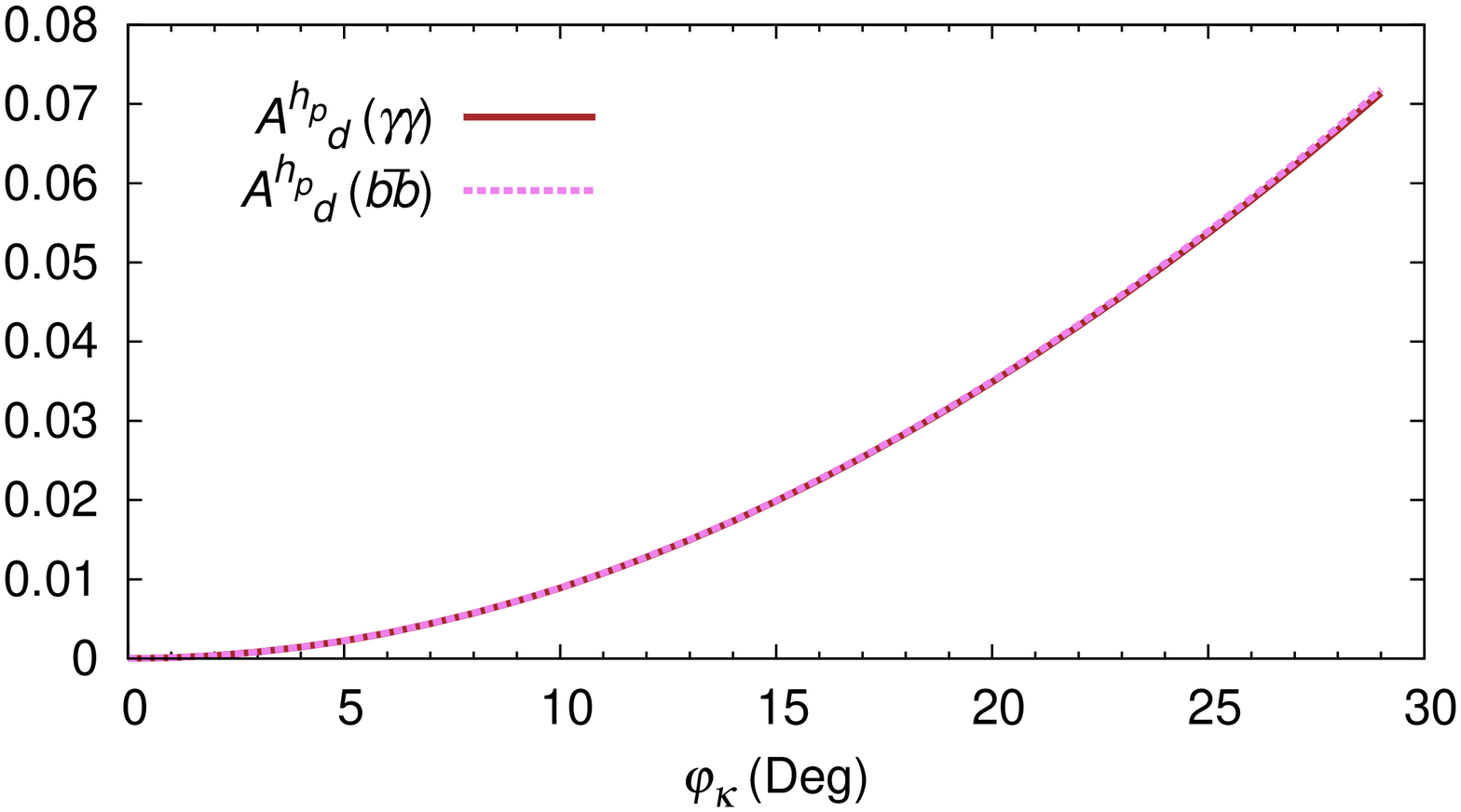}
}
\subfloat[]{%
\includegraphics*[angle=-90,scale=0.285]{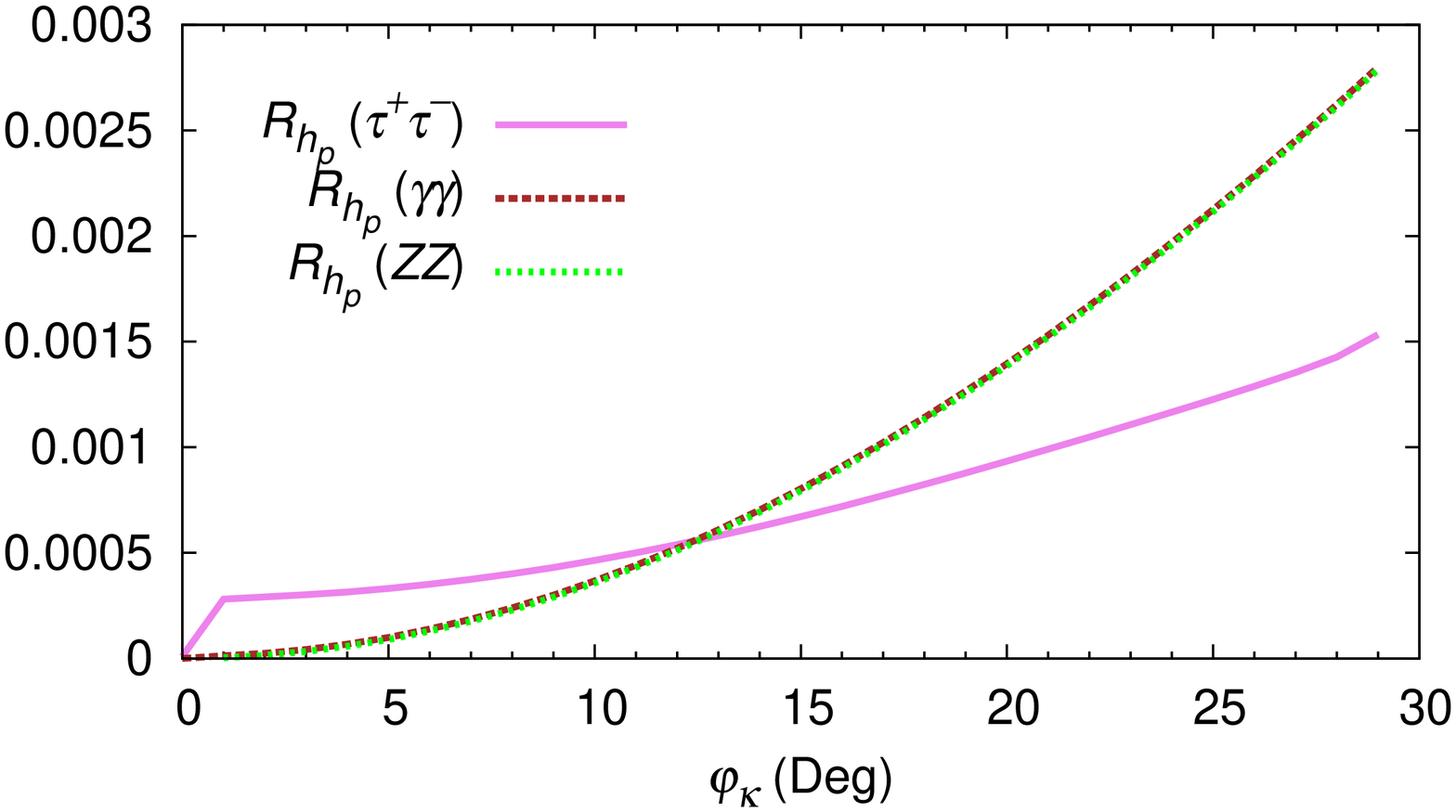}
}

\centering
\subfloat[]{%
\includegraphics*[angle=-90,scale=0.285]{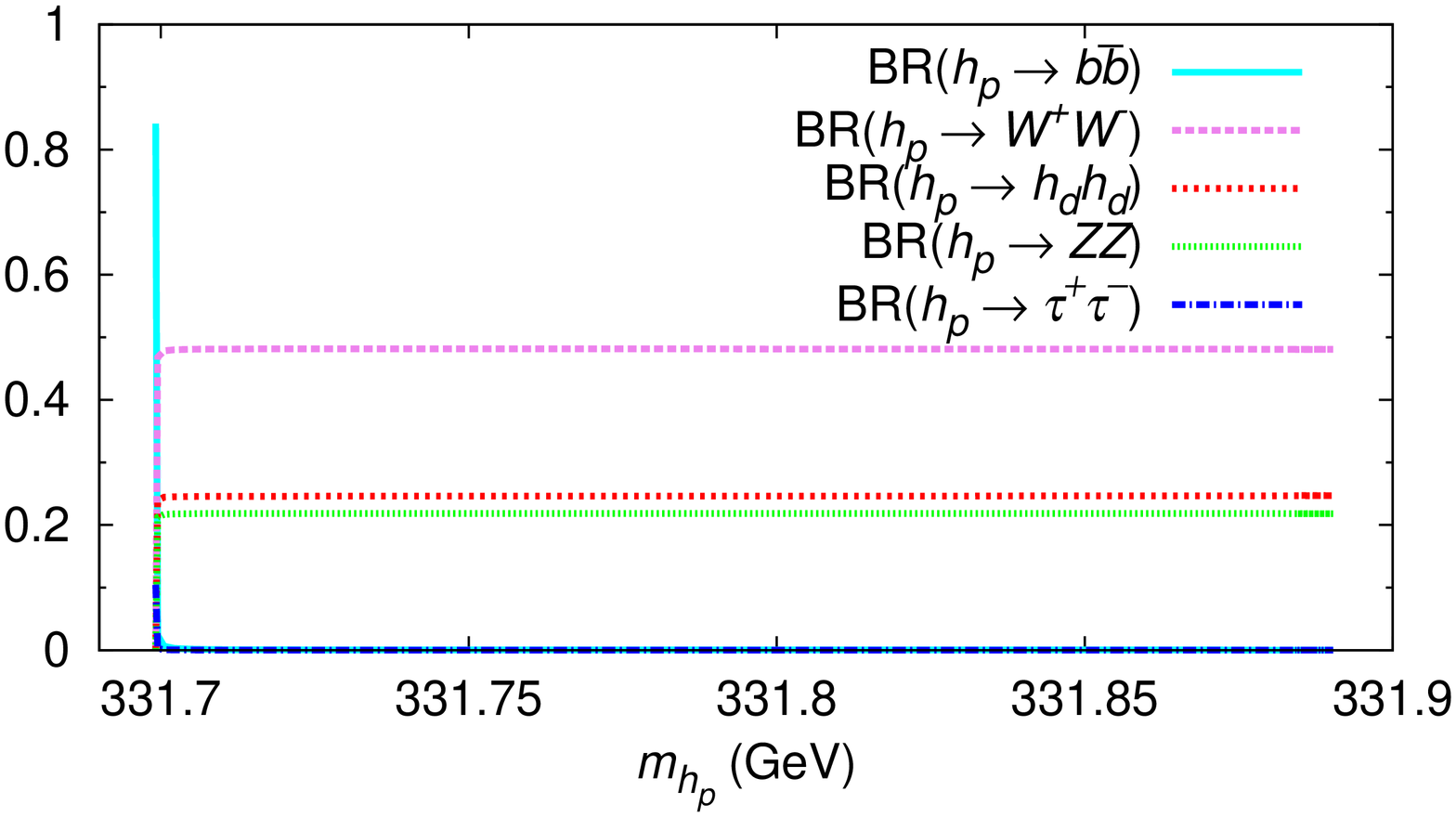}
}
\subfloat[]{%
\includegraphics*[angle=-90,scale=0.285]{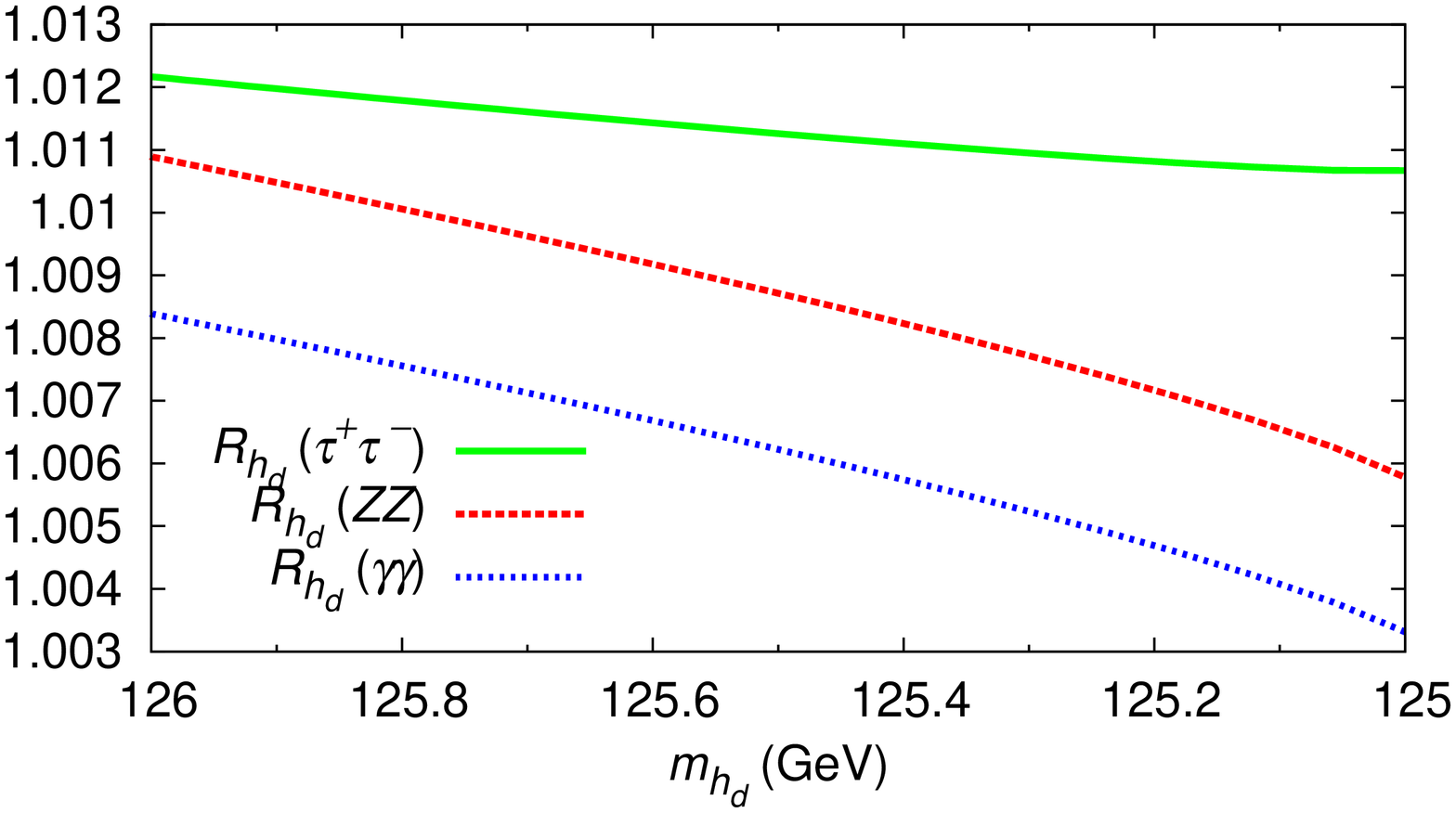}
}
\caption{Case when $h_1 = h_d$ and $h_2 = h_p$. (a) Auxiliary rates
  $A^{h_p}_{d}(\gamma\gamma)$ (solid brown line) and $A^{h_p}_{d}(b\bar{b})$
(dashed violet line) as functions of \phkap, for $h_d h_d$ pair production. (b) 
$R_{h_p}(\tau^+\tau^-)$ (solid violet line), $R_{h_p}(\gamma\gamma)$
(dashed brown line) and $R_{h_p}(ZZ)$ (dotted green line) as functions of \phkap. (c) BRs of $h_p$ into $b\bar{b}$ (solid cyan line), $W^+W^-$ (dashed violet line), $h_dh_d$
(large-dotted red line), $ZZ$ (small-dotted green line) and
$\tau^+\tau^-$ (dot-dashed blue line) vs $m_{h_p}$. (d) Signal strengths of $h_d$ in the
$\tau^+\tau^-$ channel (solid green line), in the $ZZ$ channel (dashed
red line) and in the $\gamma\gamma$ channel
(dotted blue line) vs $m_{h_d}$.}
\label{fig:C1}
\end{figure}

We first discuss the case when the lightest Higgs state, $h_1$, is
SM-like while $h_p$ is the second lightest of the five neutral Higgs
states of the model, hence corresponding to $h_2$. As a representative
of this case we choose the point P1, given in table\,\ref{tab:RPs}, 
in the cNMSSM parameter space.
\begin{table}[t]
\begin{center}
\begin{tabular}{|c|c|c|c|c|c|c|c|c|c|}
\hline
Point &$M_0$ & $M_{1/2}$ & $A_0$ & $\tan\beta$ & $\lambda$ & $\kappa$ &
$\mu_{\rm eff}$ & $A_\lambda$ & $A_\kappa$ \\ 
\hline  
\hline
P1 & 2500 & 1300 & -6000 & 12 & 0.09 & 0.11 & 1000 &
600 & -30 \\ 
\hline
P2 & 2500 & 1000 & -3000 & 20 & 0.04 & 0.013 & 200
& 200 & -200 \\ 
\hline
P3 & 1000 & 500 & -2500 & 2 & 0.54 & 0.34 & 140 & 185 & -200 \\
\hline
\end{tabular} 
\caption{\label{tab:RPs} Values of the cNMSSM parameters corresponding to
  the three benchmark cases discussed in the text. All dimensionful
  parameters are in units of GeV.}
\end{center}
\end{table}
This point yields $h_d$ around 125\,GeV in the CPC limit, with
almost exactly SM-like signal strengths in the $\gamma\gamma$, $ZZ$
and $\tau^+\tau^-$ channels,
despite a non-vanishing $\lambda$ and, hence, singlet component (such
a NMSSM Higgs boson has been discussed in\,\cite{Badziak:2013bda}). 
In panel (a) of figure\,\ref{fig:C1} we show the
auxiliary signal rates $A^{h_p}_{d}(\gamma\gamma)$ and $A^{h_p}_{d}(b\bar{b})$ as functions of
\phkap\ for P1. We see that the lines corresponding to these two signal rates overlap each other
exactly. Both these rates rise
gradually and reach a maximum value, $\sim 0.07$, for $\phkap =
29^\circ$. Such a $h_p$ can thus be responsible for up to 7\% of the
observed $\gamma\gamma$
excess besides that due to the direct production of $h_d$ in the gluon
fusion channel. The increase in $A^{h_p}_{d}(\gamma\gamma)$ and $A^{h_p}_{d}(b\bar{b})$ with
\phkap\ is a twofold consequence of the gradual increase in the
gluonic width of $h_p$ and an increase in its BR into the $h_d$ pair. The
reason for the cutoff in the line is that beyond $\phkap = 29^\circ$ 
the minimization condition given in Eq.\,(\ref{eq:ilik}) is not satisfied any more.

In panel (b) we show the signal strength of $h_p$, produced
via gluon fusion, in the $\gamma\gamma$, $ZZ$ and $\tau^+\tau^-$decay
channels. We note that although
there is a considerable rise in $R_{h_p}$, particularly in the
$\gamma\gamma$ and $ZZ$ channels, with
an increasing amount of CP violation, these rates 
barely exceed the per mil level for allowed
values of \phkap. This is due to the fact that $h_p$ has a
significantly reduced coupling to two photons compared to that of a SM
Higgs boson with the same mass. In panel (c) there are shown the dominant
BRs of $h_p$ against its mass, with \phkap\ increasing from left to
right. This plot demonstrates the main reason of large auxiliary
signal rates of $h_p$ for non-zero \phkap, as observed above. We see
that as soon as the process $h_p \rightarrow h_dh_d$ is allowed, it
becomes one of the dominant decay modes of $h_p$, with BR reaching up
to $\sim 0.23$. However, it is still not the most dominant decay
mode due to the fact that $h_p$ develops non-zero couplings
also to gauge boson pairs. Therefore, the decay $h_p \rightarrow
W^+W^-$ has the highest BR for non-zero \phkap,
while the BR of $h_p$ into $ZZ$ also lies
close to that into $h_dh_d$. As a result, the decay modes
$h_p\rightarrow b\bar{b}$ and $h_p\rightarrow \tau^+\tau^-$, which had
the highest and second highest BRs, respectively, in the CPC limit, become very subdominant. Since there is a negligible
increase in the mass of $h_p$ with increasing \phkap,
all the above BRs remain almost constant over the entire allowed range of this phase. 

Finally, in panel (d) we show the signal strengths of $h_d$ in the $\gamma\gamma$,
$ZZ$ and $\tau^+\tau^-$ channels
plotted against its mass. With increasing \phkap\ (again, from left to
right) $m_{h_d}$ falls slowly. It reaches $\sim 125$\,GeV for
$\phkap = 29^\circ$, hence becoming more
consistent with the mass measurements
at the LHC\,\cite{CMS-PAS-HIG-13-005,ATLAS-CONF-2013-034} 
(which, nevertheless have appreciable experimental errors). We see in
the figure that the signal strengths of $h_d$ in all three decay modes considered
are very SM-like in the CPC limit and show a very slow drop with
increasing \phkap.

\subsection{$h_2 = h_d$}

As stated in the Introduction, in the NMSSM the $h_2$ (the second lightest scalar
in the CPC limit) can also be the $\sim 125$\,GeV SM-like Higgs boson
with the $h_1$ corresponding to $h_s$. Below, we discuss two distinct cases,
based on the compositions of $h_1$ and $h_2$, in which this
possibility is realised. \\

\noindent \underline{Small singlet-doublet mixing}: For small
$\lambda$, $\kappa$ and $\mu_{\rm eff}$ but intermediate-to-large
$\tan\beta$, $h_2$ is still doublet-dominated and hence possesses
very SM-like couplings to fermions and bosons. In this case, due to a
smaller VeV $s$ resulting from a lower value of $\mu_{\rm eff}$ (recall
that $\mu_{\rm eff} = \lambda s$) compared to the case discussed
above, the mass of the singlet-like scalar Higgs boson falls below
that of $h_d$. In fact, owing to a highly dominant
singlet component, $m_{h_s}$ can reach very low values, $\sim
40$\,GeV, before it violates the LEP limit
on $hZ$ production\,\cite{Barate:2003sz}.
This effectively bounds $m_{h_p}$, which grows
with increasing $A_\kappa$ while $m_{h_s}$ falls, from above. Thus it is extremely
difficult for $A_\kappa$ and, resultantly,
$m_{h_p}$ to become large enough to allow the  $h_p \rightarrow h_dh_d$
decay. However, thanks to a fairly light $h_s$, the decay $h_p
\rightarrow h_dh_s$ is alternatively possible for non-zero \phkap. 

\begin{figure}[t]
\centering
\subfloat[]{%
\includegraphics*[angle=-90,scale=0.285]{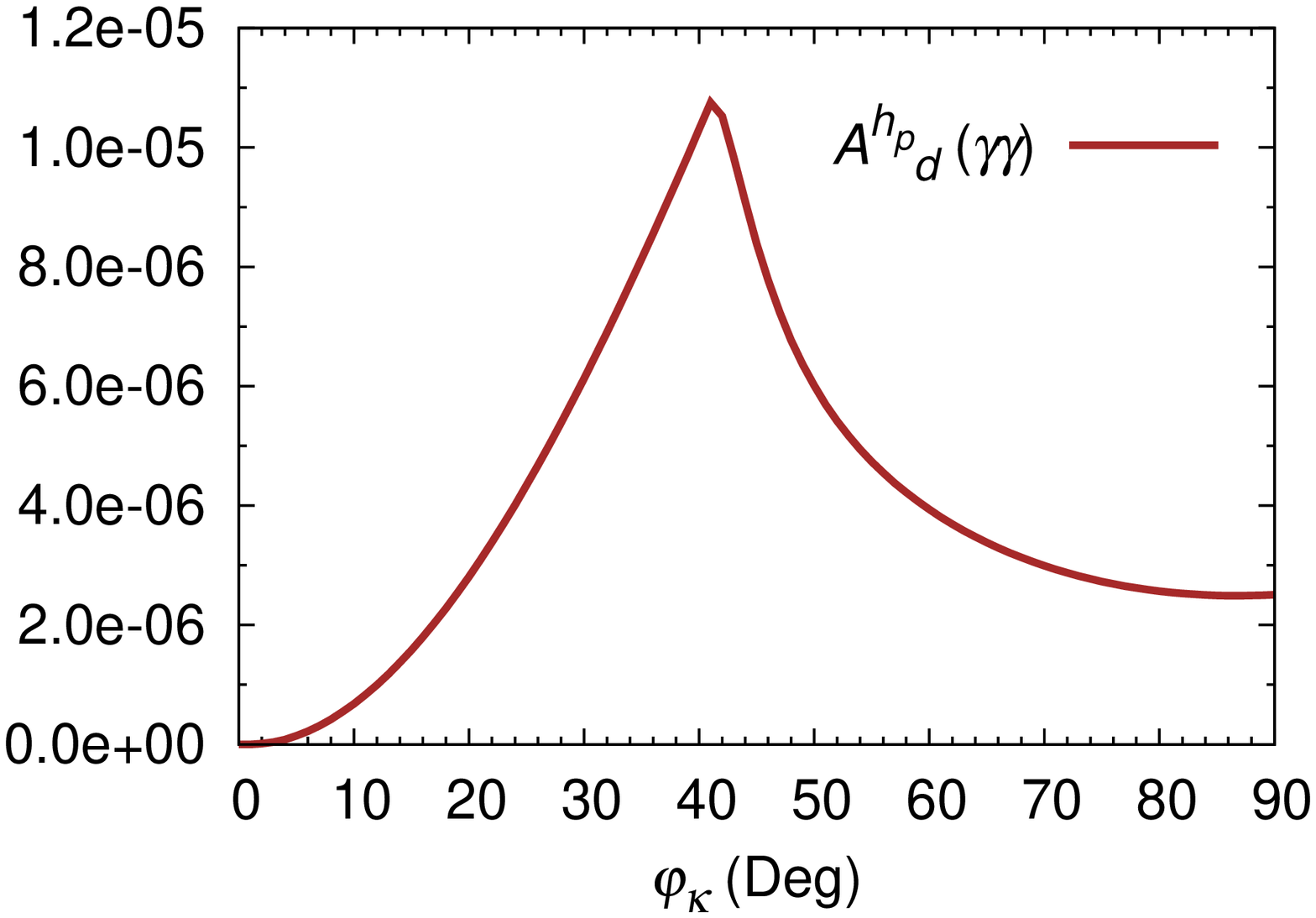}
}
\subfloat[]{%
\hspace*{-0.5cm}\includegraphics*[angle=-90,scale=0.285]{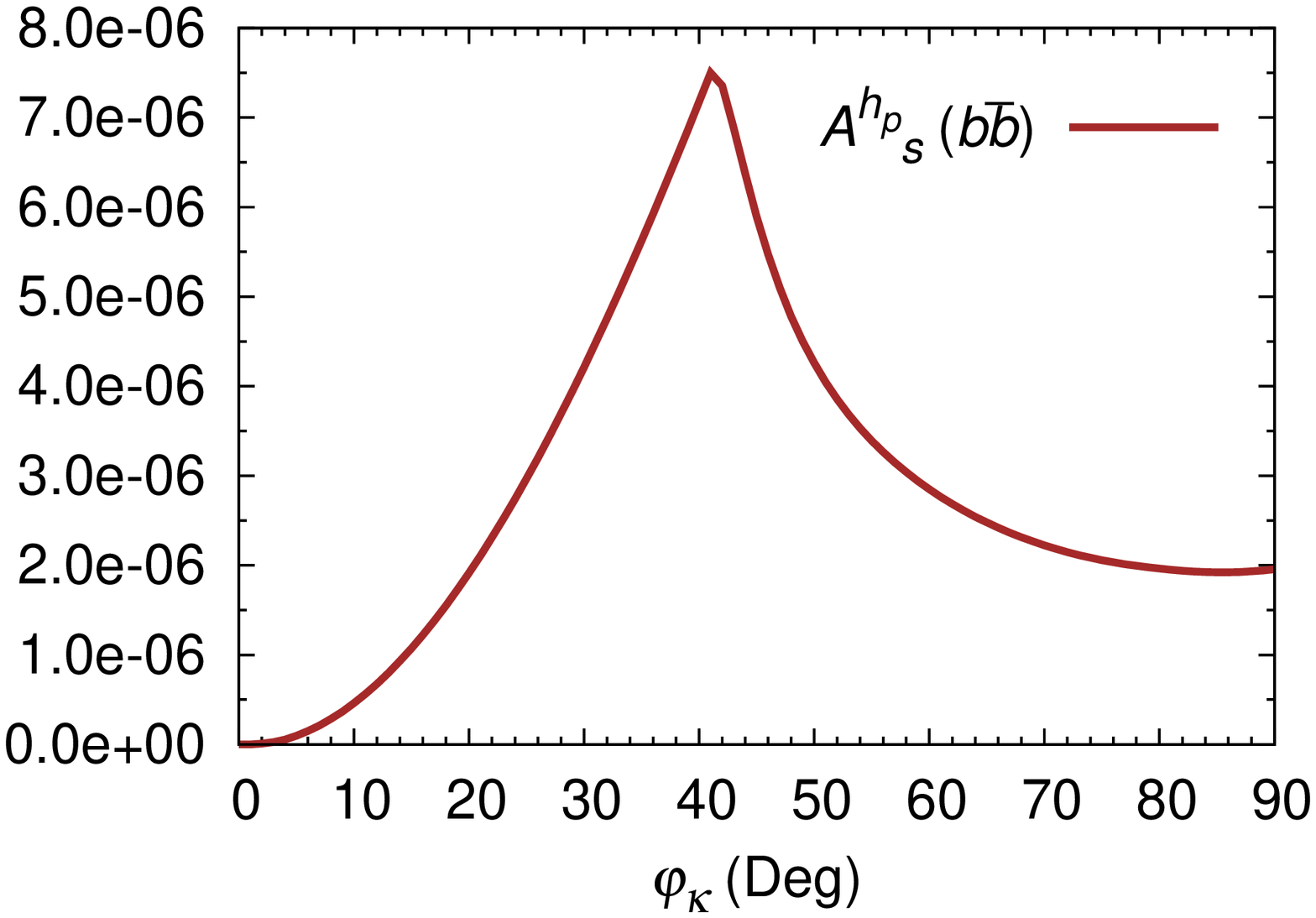}
}
\subfloat[]{%
\hspace*{-0.5cm}\includegraphics*[angle=-90,scale=0.285]{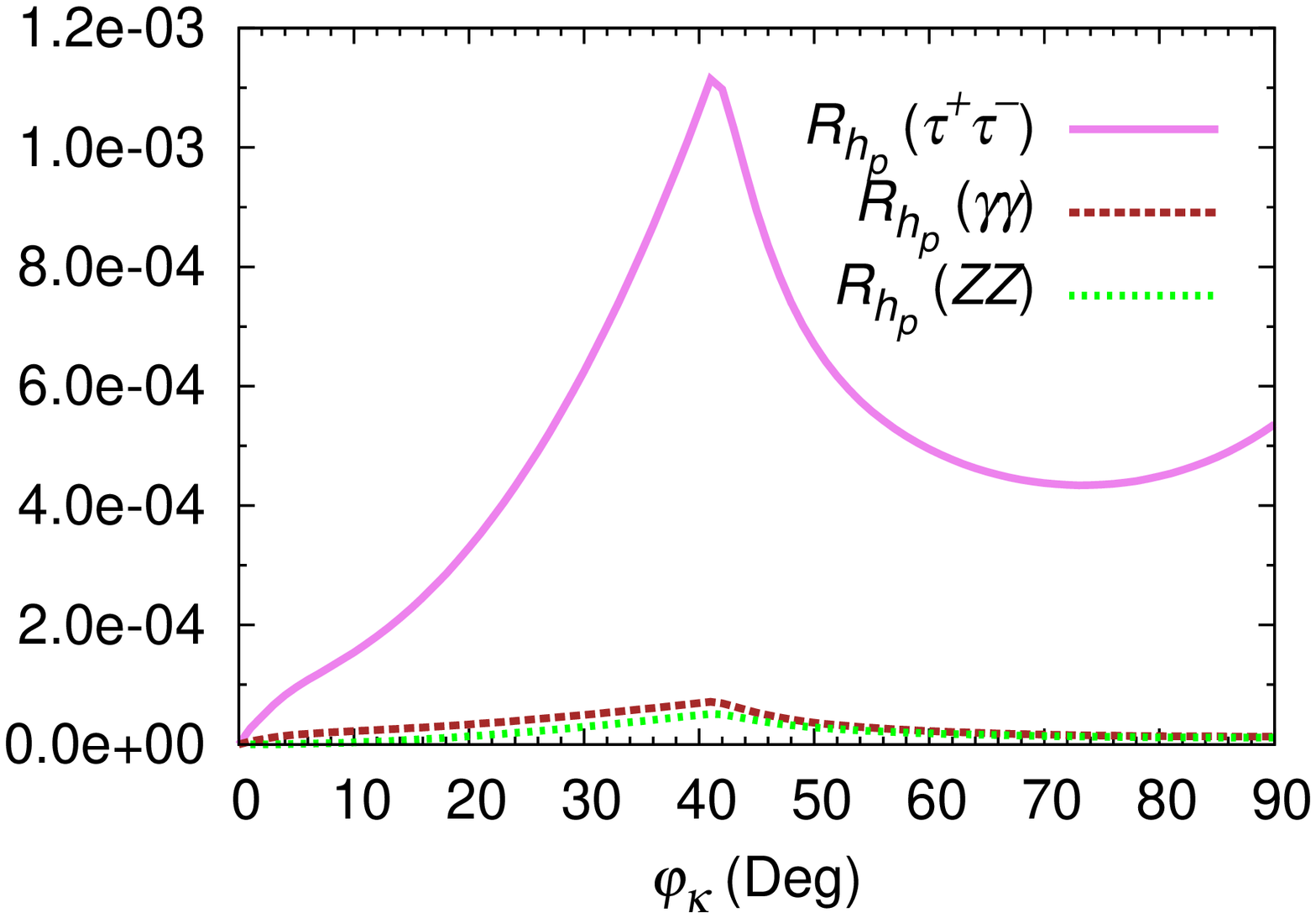}
}

\centering
\subfloat[]{%
\includegraphics*[angle=-90,scale=0.285]{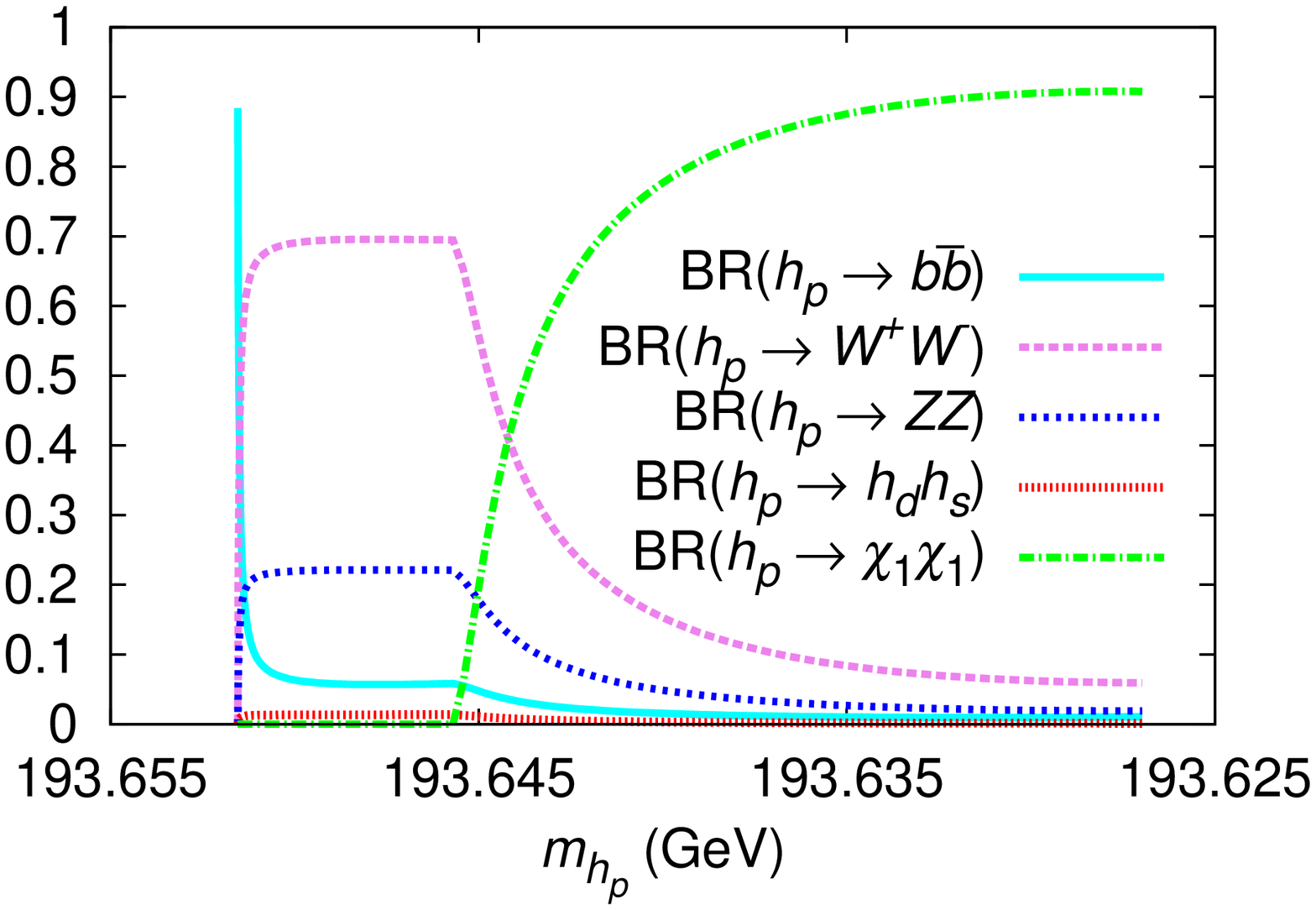}
}
\subfloat[]{%
\hspace*{-0.5cm}\includegraphics*[angle=-90,scale=0.285]{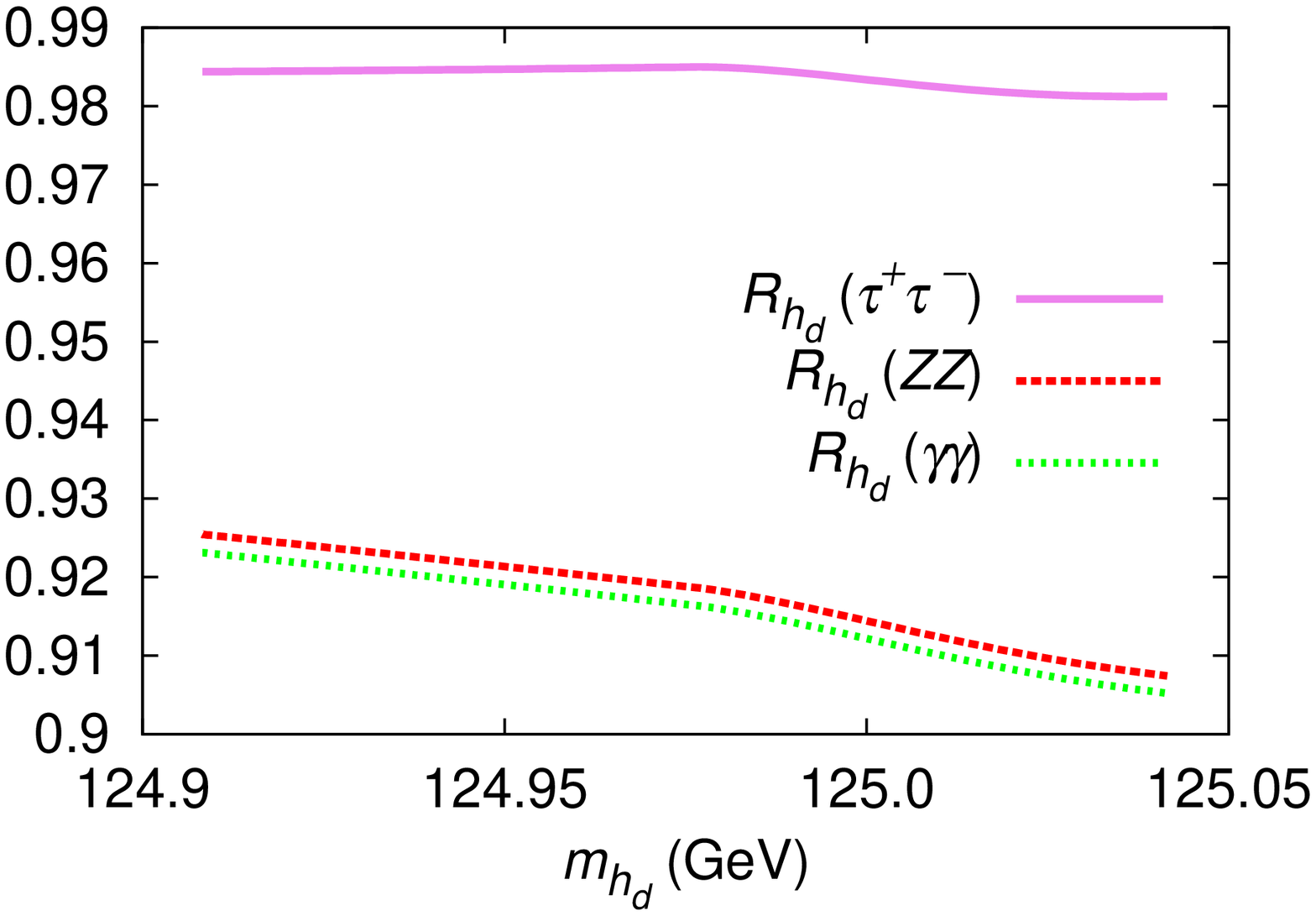}
}
\subfloat[]{%
\hspace*{-0.5cm}\includegraphics*[angle=-90,scale=0.285]{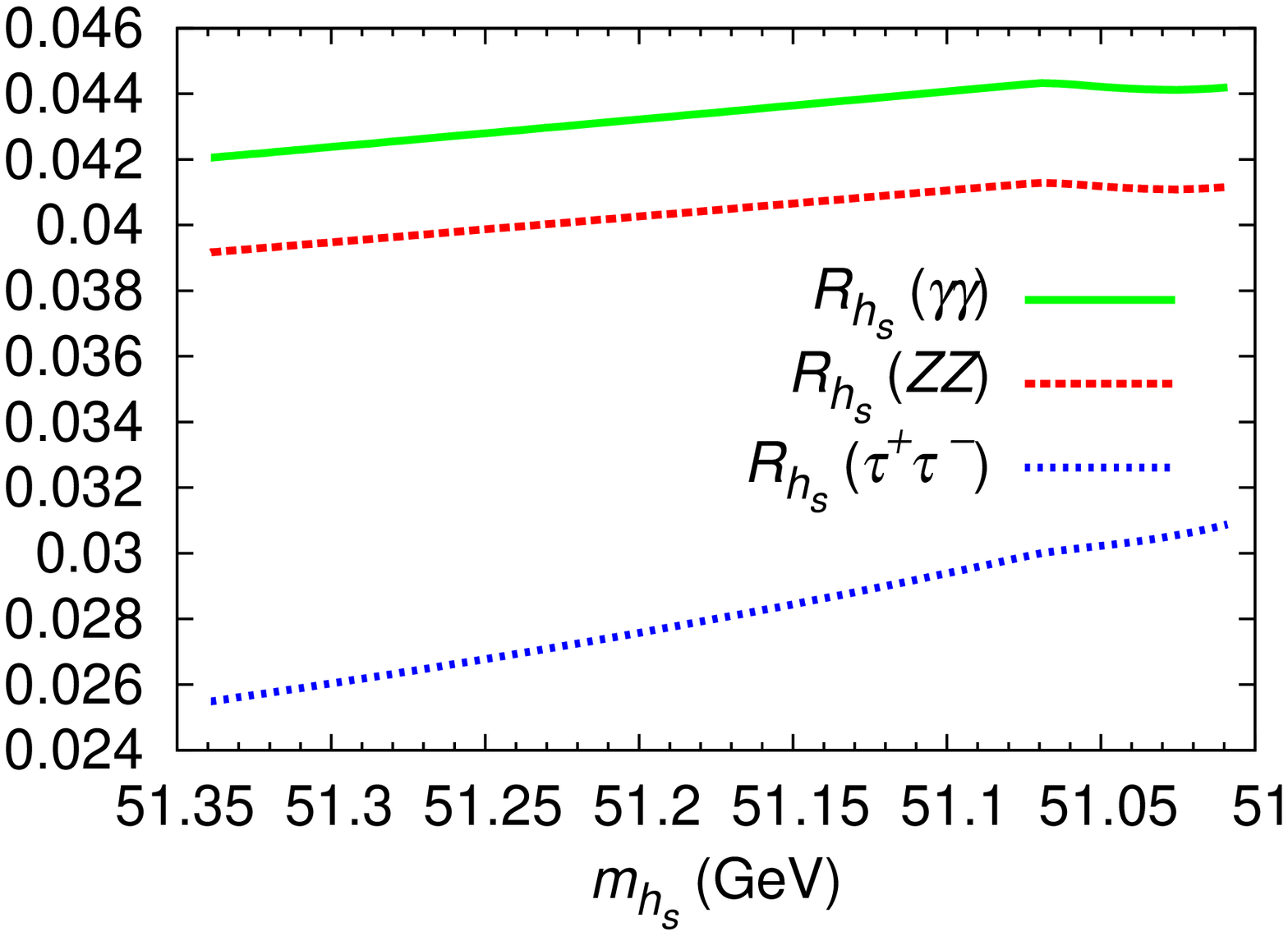}
}
\caption{Case when $h_2 = h_d$ with small singlet-doublet mixing and
  $h_3 = h_p$.  (a),\,(b) Auxiliary rates
  $A^{h_p}_{d}(\gamma\gamma)$ and $A^{h_p}_{s}(b\bar{b})$, respectively, as
  functions of \phkap, for $h_d h_s$ pair production. (c) 
$R_{h_p}(\tau^+\tau^-)$ (solid violet line), $R_{h_p}(\gamma\gamma)$
(dashed brown line) and $R_{h_p}(ZZ)$ (dotted green line) as functions of \phkap. (d) BRs of $h_p$ into 
$b\bar{b}$ (solid cyan line), $W^+W^-$ (dashed violet line), $ZZ$
(large-dotted blue line), $h_dh_s$ (small-dotted red line) and
$\chi_1\chi_1$ (dot-dashed green line) vs $m_{h_p}$. (e) Signal strengths of $h_d$ in the
$\tau^+\tau^-$ channel (solid violet line), in the $ZZ$ channel (dashed
red line) and in the $\gamma\gamma$ channel
(dotted green line) vs $m_{h_d}$. (f) Signal strengths of $h_s$ in the
$\gamma\gamma$ channel (solid green line), in the $ZZ$ channel (dashed
red line) and in the $\tau^+\tau^-$ channel (dotted blue line) vs $m_{h_s}$.}
\label{fig:C2}
\end{figure}

We choose the point P2, with its coordinates in
the cNMSSM parameter space given in table\,\ref{tab:RPs}, to
demonstrate the effects of CP violation on the phenomenology of $h_p$ for this
case. In panel (a) of figure\,\ref{fig:C2} we
show $A^{h_p}_{d}(\gamma\gamma)$ against \phkap\ for P2. We see in
the figure that $A^{h_p}_{d}(\gamma\gamma)$ grows steadily until
$\phkap=40^\circ$ after which it falls abruptly. The reason for this
fall is the opening up of the $h_p \rightarrow \chi_1\chi_1$ decay
channel as we shall see below. Note that even the peak value of
$A^{h_p}_{d}(\gamma\gamma)$ for $\phkap=40^\circ$ in this case lies two
orders of magnitude below the per mil level. The line has been
artificially cut off at
$\phkap=90^\circ$ since the auxiliary rate remains almost steady
afterwards. In panel (b)
$A^{h_p}_{s}(b\bar{b})$ is shown for the second Higgs boson, $h_s$,
produced along with $h_d$ against \phkap. The auxiliary rate via this Higgs
boson is always lower
than that of $h_d$ on account of its being singlet-dominated and hence
coupling very weakly to matter. In panel (c) we show the direct
production signal rates of $h_p$ in the $\gamma\gamma$, $ZZ$ and
$\tau^+\tau^-$ channels against \phkap. While $R_{h_p} (\gamma\gamma)$ and $R_{h_p}
(ZZ)$ remain almost of the same order as the auxiliary rate via $h_d$, $R_{h_p}
(\tau^+\tau^-)$ rises much more briskly with increasing \phkap\ and
reaches the per mil level for $\phkap\sim 40^\circ$. 

The reason for the sudden drop in the various signal rates of $h_p$
after $\phkap=40^\circ$ becomes obvious from panel (d), where we show
its dominant BRs plotted against $m_{h_p}$. In contrast with the first
case above, even when the $h_p \rightarrow h_d h_s$ decay channel 
opens up for non-zero \phkap, it remains very subdominant, with BR still
smaller than that for the $h_p \rightarrow b\bar{b}$ mode. We see in the figure that for small non-zero values of \phkap\ the decay mode $h_p
\rightarrow W^+W^-$ is clearly the most dominant one, with
BR as high as $\sim 0.7$, while $h_p \rightarrow ZZ$ is the second most dominant mode. With increasing \phkap\
(left to right) $m_{h_p}$ falls negligibly, but just before it reaches
187.86\,GeV, the BR($h_p\rightarrow \chi_1 \chi_1$) suddenly shoots up. 
This is a consequence of the fact that $m_{\chi_1}$ also falls
sharply as \phkap\ is increased, so much so that for $\phkap >40^\circ$
$\chi_1$ becomes light enough to make the decay of $h_p$ into its pair possible kinematically.
Resultantly, beyond $\phkap =40^\circ$ all the hitherto dominant decay
modes, $h_p \rightarrow W^+W^-$, $h_p \rightarrow ZZ$ and $h_p
\rightarrow b\bar{b}$, become more and more subdominant while the BR($h_p \rightarrow h_dh_s$) falls even further.

The above discussion of the behaviour of various BRs of $h_p$ has an
important implication, that $\chi_1$, at least for large
values of \phkap, is highly singlino-dominated. It should, therefore, be
extremely difficult to be probed at a direct detection experiment for dark matter, such as
XENON\,\cite{Aprile:2012nq}. In panel (e) we show the signal strengths
of $h_d$ against its mass for this case. The mass $m_{h_d}$ increases slowly
with increasing \phkap, conversely to P1, while $R_{h_d}(ZZ)$ and
$R_{h_d}(\gamma\gamma)$ fall gradually. These two rates
never drop below 0.9 and hence always lie well within the 
experimental uncertainties around the measured central values at the
LHC. $R_{h_d}(\tau^+\tau^-)$, on
the other hand, is always much higher than $R_{h_d}(ZZ)$ and
$R_{h_d}(\gamma\gamma)$ and closer to 1 for all values of
\phkap. Finally, in panel (f) there are shown the signal strengths of the
accompanying $h_s$ in the same three decay channels. Conversely to the
$h_d$ rates, $R_{h_s}(\tau^+\tau^-)$ is much lower than $R_{h_s}(ZZ)$ and
$R_{h_s}(\gamma\gamma)$, with all these rates lying just above the
percent level for $\phkap=0^\circ$. The rates rise slowly with
increasing \phkap\ until it reaches $40^\circ$, after which they
become almost steady. \\

\noindent \underline{Large singlet-doublet mixing}: It was noted in
\cite{Ellwanger:2011aa} that, for large $\lambda$ and $\kappa$ and small
$\tan\beta$ and $\mu_{\rm eff}$, $h_2$ in the NMSSM (again, $h_d$ here) can have a considerably enhanced $\gamma\gamma$ rate
compared to $h_{\rm SM}$, due mainly to the reduced coupling and
consequently reduced BR($h_d \rightarrow b\bar{b}$). This scenario, in
which the $h_1$ ($h_s$ here) has a mass lying just below
$m_{h_2}$\,\cite{Gunion:2012gc} and the lightest stop can have 
a mass significantly below 1\,TeV\,\cite{Ellwanger:2012ke}, is sometimes
referred to as the `natural NMSSM'\,\cite{Kang:2013rj}. To discuss the
impact of a light $h_p$ on such a scenario in the cNMSSM we choose the
point P3, given in table\,\ref{tab:RPs}.

\begin{figure}[t]
\centering
\subfloat[]{%
\includegraphics*[angle=-90,scale=0.285]{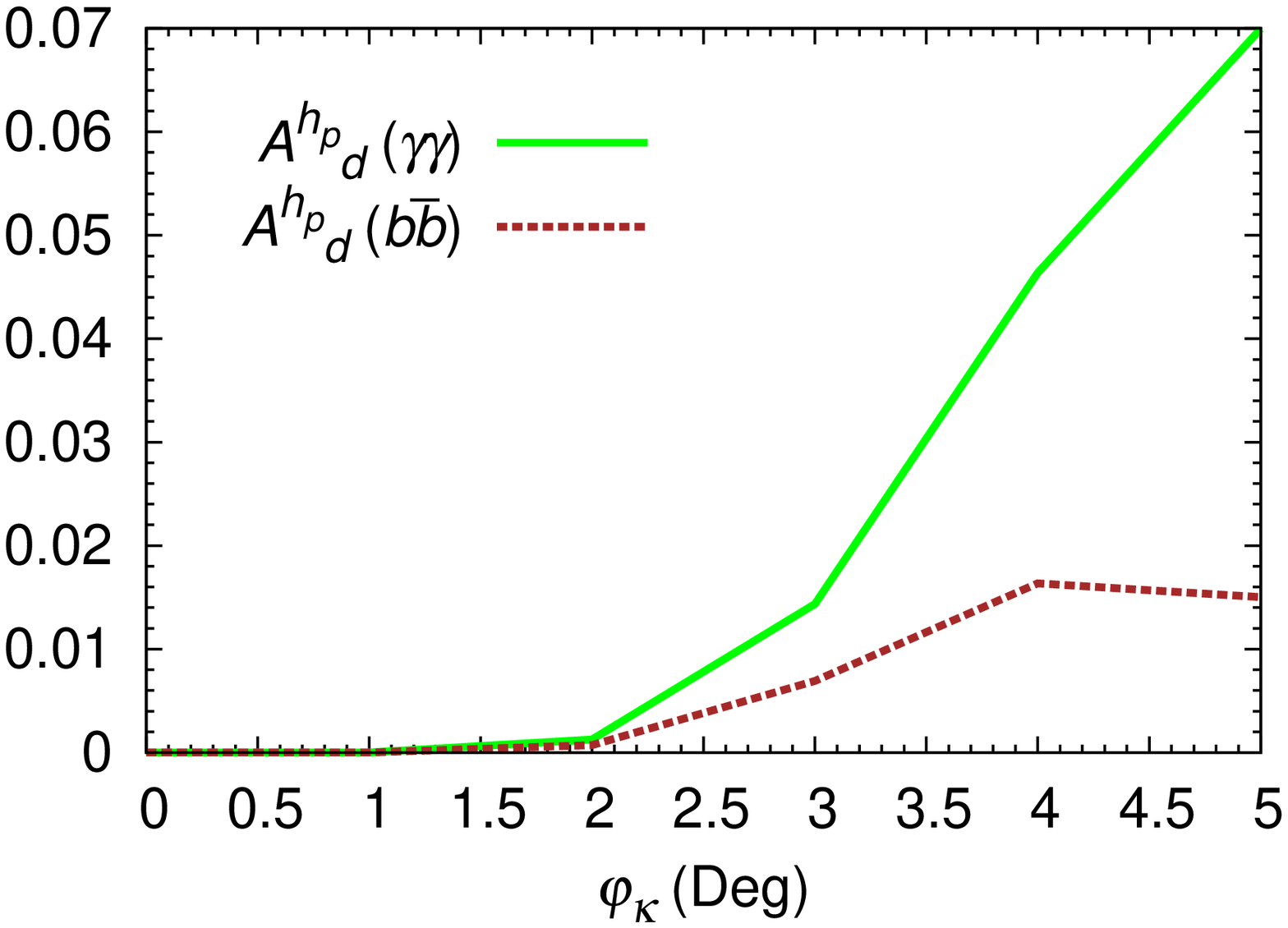}
}
\subfloat[]{%
\hspace*{-0.5cm}\includegraphics*[angle=-90,scale=0.285]{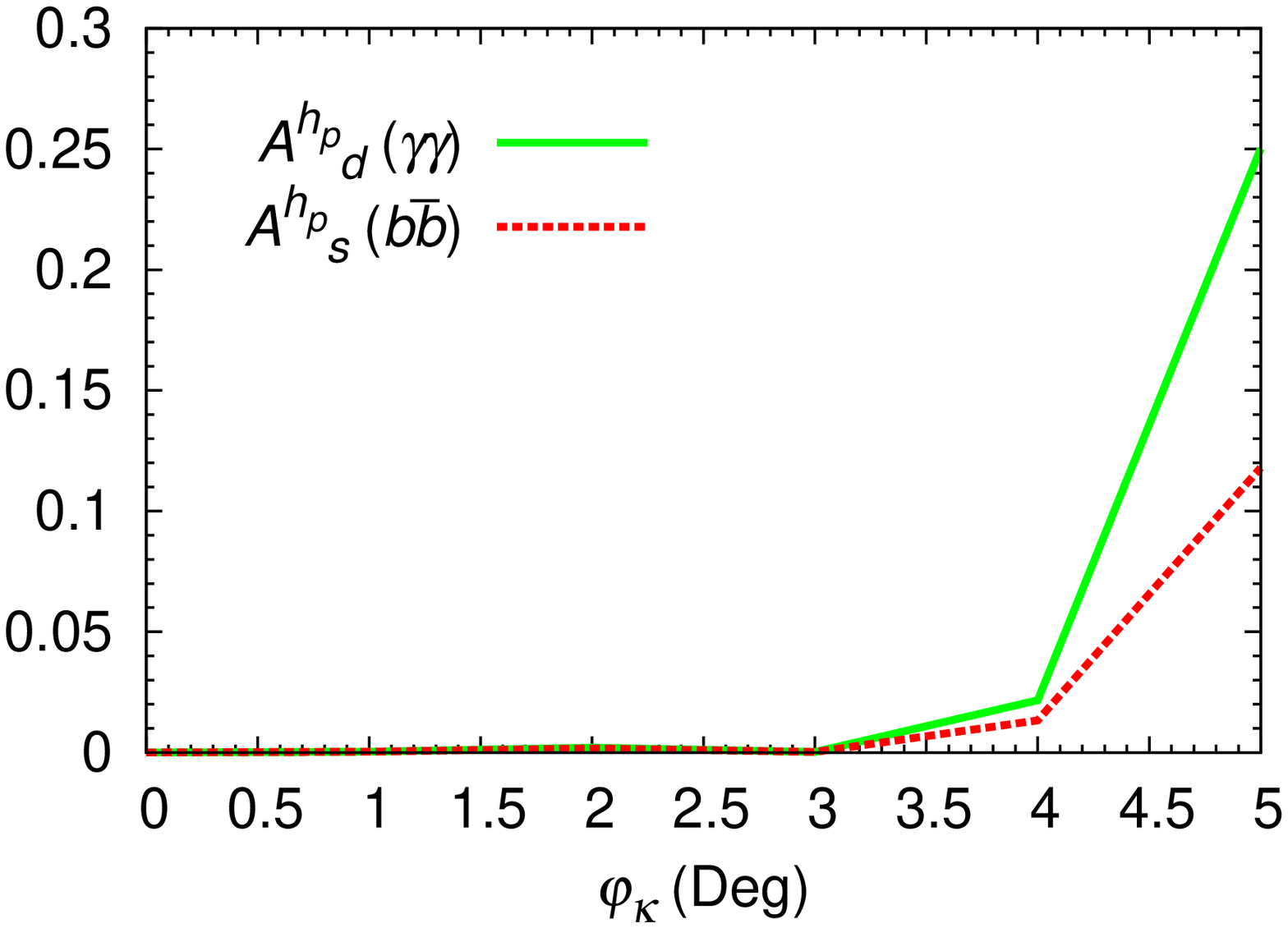}
}
\subfloat[]{%
\hspace*{-0.5cm}\includegraphics*[angle=-90,scale=0.285]{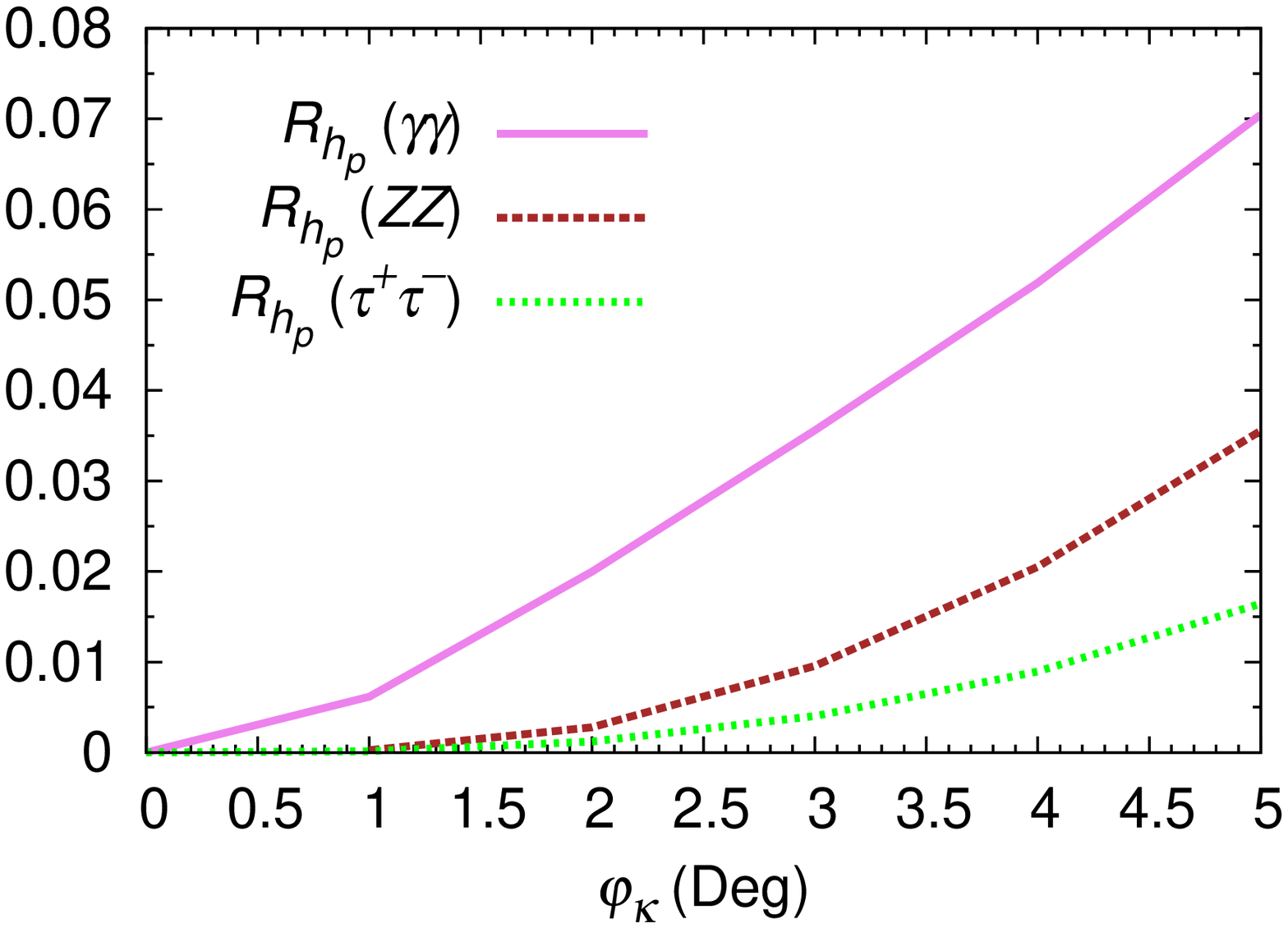}
}

\centering
\subfloat[]{%
\includegraphics*[angle=-90,scale=0.285]{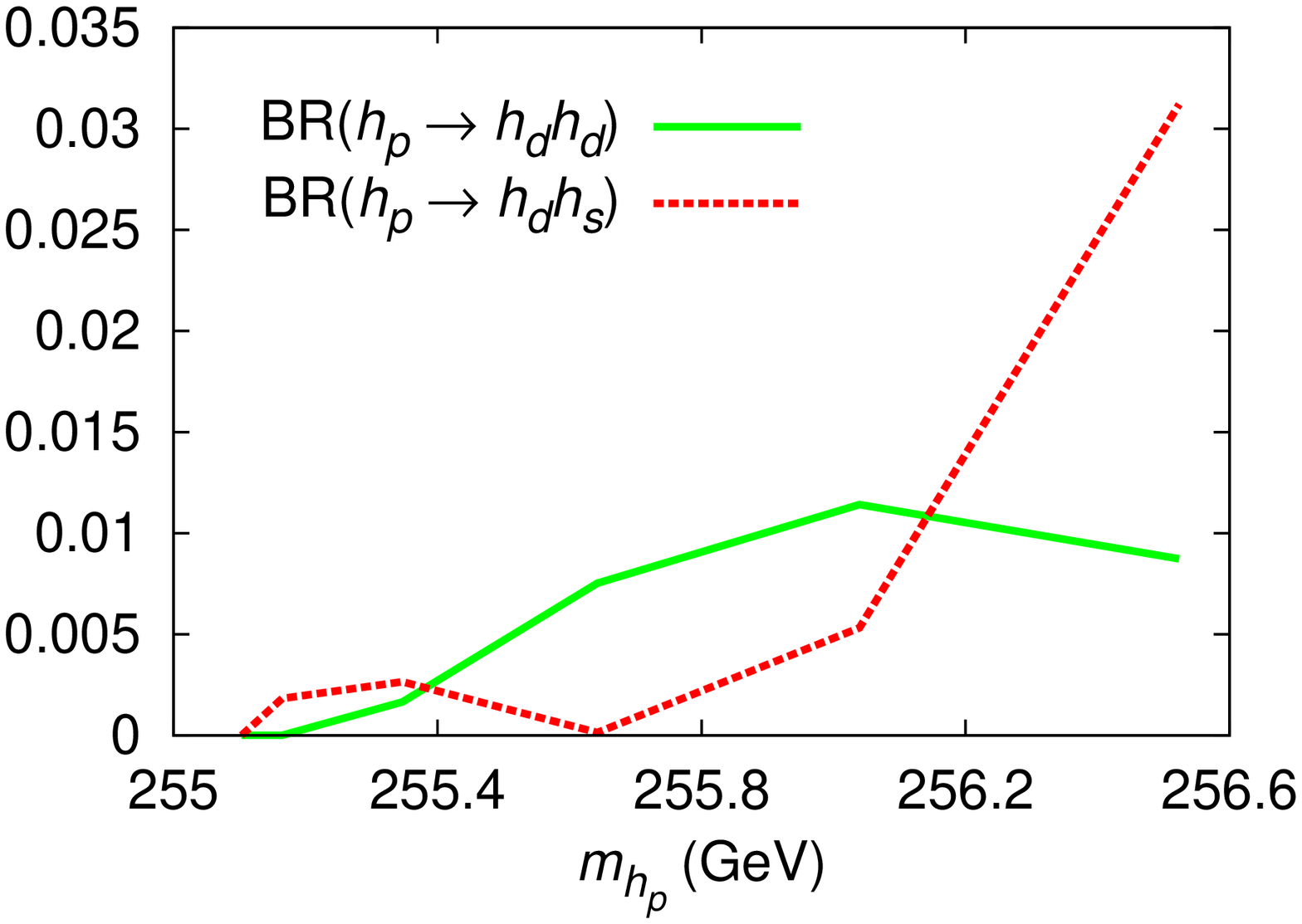}
}
\subfloat[]{%
\hspace*{-0.5cm}\includegraphics*[angle=-90,scale=0.285]{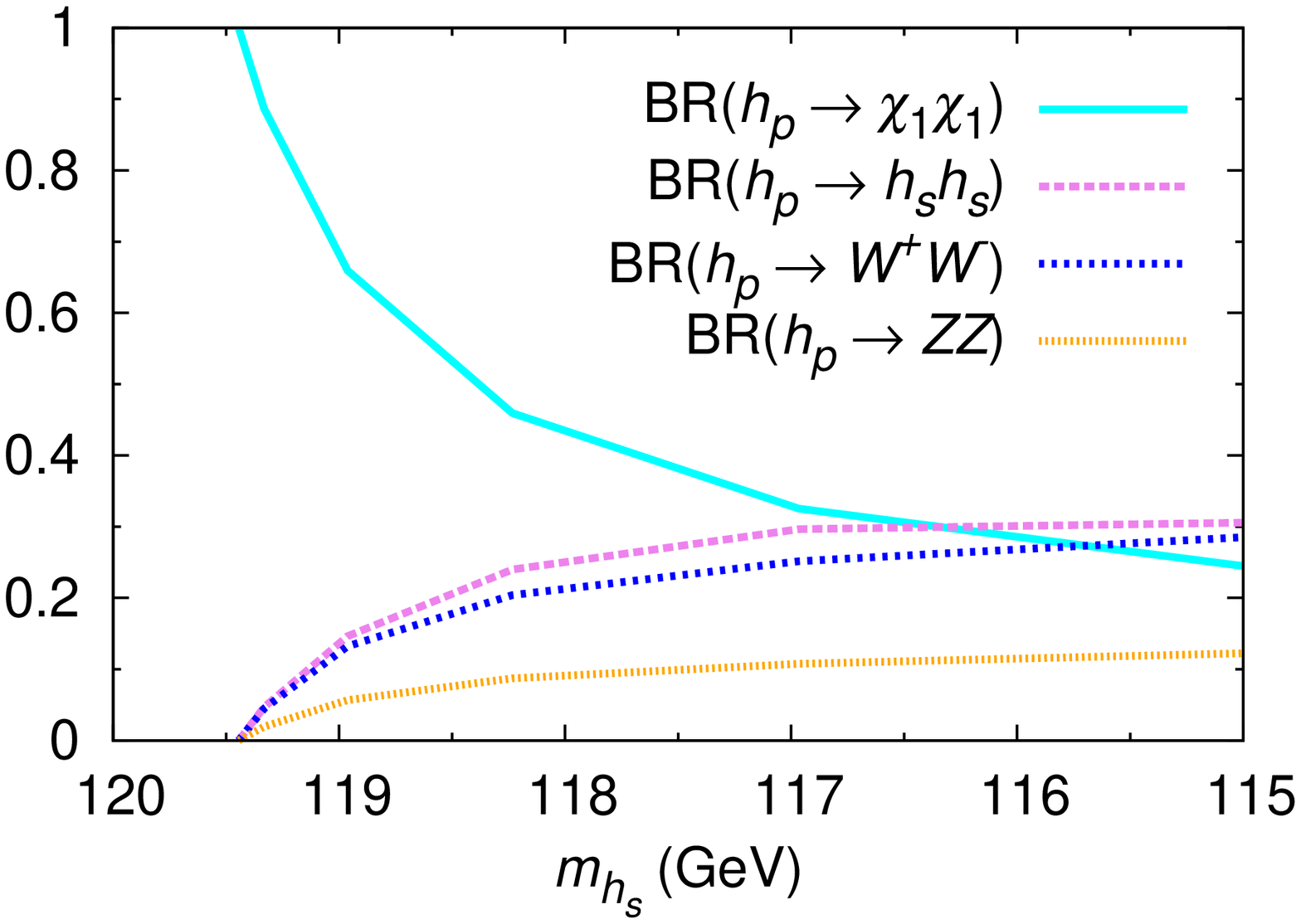}
}
\subfloat[]{%
\hspace*{-0.5cm}\includegraphics*[angle=-90,scale=0.285]{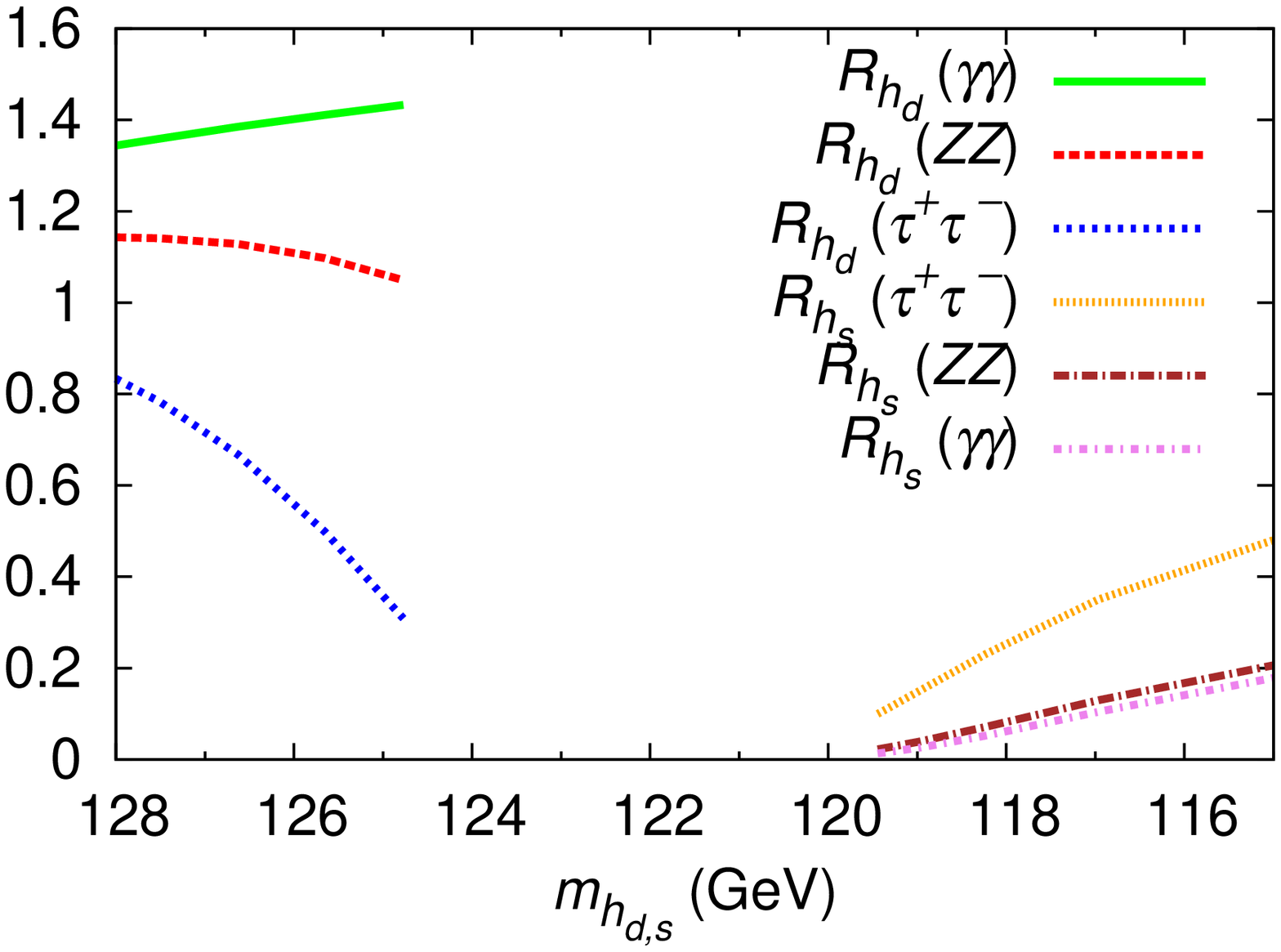}
}
\caption{Case when $h_2 = h_d$ with large singlet-doublet mixing and
  $h_3 = h_p$.
(a) Auxiliary rates $A^{h_p}_{d}(\gamma\gamma)$ (solid green line) and
 $A^{h_p}_{d}(b\bar{b})$ (dashed brown line) as functions of \phkap,
 for $h_d h_d$ pair production. (b)
  $A^{h_p}_{d}(\gamma\gamma)$ (solid green line) and
  $A^{h_p}_{s}(b\bar{b})$ (dashed red line) as
  functions of \phkap, for $h_d h_s$ pair production. (c) 
Signal strengths $R_{h_p}(\gamma\gamma)$ (solid violet line), $R_{h_p}(ZZ)$
(dashed brown line) and $R_{h_p}(\tau^+\tau^-)$ (dotted green line) as
functions of \phkap. (d) BRs of $h_p$ into 
$h_dh_d$ (solid green line) and $h_dh_s$ (dashed red line) vs 
$m_{h_p}$. (e) BRs of $h_p$ into $\chi_1\chi_1$ (solid cyan line),
$h_sh_s$ (dashed violet line), $W^+W^-$ (large-dotted blue line) and
$ZZ$ (small-dotted orange line)  vs $m_{h_s}$. (f)
$R_{h_d}(\gamma\gamma)$ (solid green line), $R_{h_d}(ZZ)$
(dashed red line), $R_{h_d}(\tau^+\tau^-)$ (large-dotted blue line), 
$R_{h_s}(\tau^+\tau^-)$ (small-dotted orange line), $R_{h_s}(ZZ)$
(dot-large-dashed brown line) and $R_{h_s}(\gamma\gamma)$
(dot-small-dashed violet line)  vs $m_{h_{d,s}}$.}
\label{fig:C3}
\end{figure}

Unlike in the second case discussed above, in this case $h_p$
can easily have a mass more than twice that of $h_d$, implying that
its decay into $h_dh_d$ is possible simultaneously with that into $h_dh_s$, 
once  CP is violated. In panel (a) of figure\,\ref{fig:C3} we show 
$A^{h_p}_{d}(\gamma\gamma)$ and
$A^{h_p}_{d}(b\bar{b})$ when a pair of $h_d$ is produced via $h_p$
decay, as functions of the
phase \phkap. We see that $A^{h_p}_{d}(\gamma\gamma)$ grows rapidly with
increasing \phkap, reaching $\sim 0.07$ for $\phkap
= 5^\circ$. $A^{h_p}_{d}(b\bar{b})$ also grows, although relatively
slowly, with increasing \phkap, which is cut off at $5^\circ$ due to the fact
that $m_{h_d}$ falls sharply, as we shall see later, and for larger
values of the phase it becomes
incompatible with the current LHC measurements of the Higgs boson
mass. At the same time, the mass of $h_s$, which has a significant
doublet component due to the large $\lambda$, also violates the LEP bound mentioned earlier. 
In panel (b) we show $A^{h_p}_{d}(\gamma\gamma)$ and
$A^{h_p}_{s}(b\bar{b})$ when, alternatively, a $h_dh_s$ pair is produced via $h_p$
decay, as functions of \phkap. In this case the two auxiliary rates
rise to much larger values for $\phkap=5^\circ$ compared to the case of $h_dh_d$ pair
production seen in panel (a). Notably, while
$A^{h_p}_{d}(\gamma\gamma)$ reaches a peak value of 0.25,
$A^{h_p}_{s}(b\bar{b})$ also rises to about 0.12, owing to the fact
that $h_s$ here has a considerably larger doublet component compared to the  
above case with small singlet-doublet mixing. 
Panel (c) shows that $R_{h_p}(\gamma\gamma)$ for this case
also rises to percent level for $\phkap>1^\circ$ and reaches a peak value
of $\sim 0.07$. $R_{h_p}(ZZ)$ and $R_{h_p}(\tau^+\tau^-)$ also
rise slowly, with the latter barely exceeding the per mil level for $\phkap=5^\circ$.

In panel (d) of figure\,\ref{fig:C3} we show the BR($h_p \rightarrow
h_dh_d$) and the BR($h_p \rightarrow h_dh_s$) 
plotted against $m_{h_p}$, with \phkap\
increasing from left to right. In contrast with the earlier cases, we
see that neither of these two BRs reaches
a value even as high as 0.04, even though they still yield significant
$A^{h_p}(\gamma\gamma)$ rates as noted above. The BR($h_p \rightarrow
h_dh_d$) is dominant over the BR($h_p \rightarrow h_dh_s$) for $\phkap \le
4^\circ$, but becomes subdominant for larger \phkap, 
owing to the fact that $m_{h_s}$ starts falling faster than $m_{h_d}$.
The reason for small BRs of $h_p$ in these two decay modes becomes
clear, once again, when one looks at the other BRs, shown in panel (e) 
against $m_{h_s}$. We see in the figure that the BR($h_p\rightarrow
\chi_1\chi_1$) is always highly dominant. In fact for $\phkap= 0^\circ$ $h_p$ almost
always decays into a pair of $\chi_1$. With increasing \phkap\
$m_{h_s}$ starts falling and, consequently, the BR($h_p\rightarrow
h_sh_s$) starts rising. At the same time the BR($h_p\rightarrow
W^+W^-$) and the BR($h_p\rightarrow ZZ$) also
rise slowly, while the BR($h_p\rightarrow \chi_1\chi_1$) drops sharply,
although it still remains the most dominant one for almost the entire allowed range
of \phkap. Only for $\phkap= 5^\circ$ the BR of hitherto the third dominant
decay mode, $h_p\rightarrow W^+W^-$, rises slightly above the BRs
of both $h_p\rightarrow \chi_1\chi_1$ and $h_p\rightarrow h_s h_s$ and 
becomes the most dominant one, $\sim 0.3$. 

Finally, in panel (f) we show the signal strengths for both $h_d$ and
$h_s$ in the $\gamma\gamma$, $ZZ$ and $\tau^+\tau^-$ channels 
against their respective masses for this case. We see that $m_{h_d}$
falls quite sharply with increasing \phkap, again in contrast
with the earlier cases, which is one of the reasons for \phkap\
being restricted to values of $\cal{O}$(1), as noted earlier. 
Additionally, $R_{h_d}(ZZ)$ is not only
smaller than $R_{h_d}(\gamma\gamma)$ when CP is
conserved but it also behaves quite differently with increasing
\phkap. $R_{h_d}(\gamma\gamma)$, already significantly above 1 in the CPC
limit, slowly increases further with increasing \phkap\ while
$R_{h_d}(ZZ)$, also slightly above 1 initially, grows more SM-like by
falling slowly. Expectedly, $R_{h_d}(\tau^+\tau^-)$ is already below
1 in the CPC limit owing to the large singlet component of $h_d$ and,
consequently, a reduced coupling to fermions. It falls further
with increasing \phkap\ and deviates considerably from a SM-like rate
for the maximum allowed value of the phase. As for $h_s$, its mass also drops with
increasing \phkap\ but its signal rates in the three decay channels
considered rise continuously. In fact, $R_{h_s}(\tau^+\tau^-)$ reaches
as high as $\sim 0.5$ for $\phkap=5^\circ$ while
$R_{h_s}(\gamma\gamma)$ and $R_{h_s}(ZZ)$ also reach up to 0.2. 

One may thus deduce in this case that non-zero values of \phkap\ are
already tightly constrained by the LHC Higgs boson data. The is due to the
dual fact that such values
push $R_{h_d}(\gamma\gamma)$, which is already on the larger side in the
CPC limit, upward, and $R_{h_d}(\tau^+\tau^-)$, which is already on
the smaller side in the CPC limit, further downward. However, it should be noted
that any further enhancement in the $\gamma\gamma$ rate is only slight
with increasing \phkap, particularly for $\phkap<4^\circ$, so that it
is still consistent with the
ATLAS measurement, $\mu(\gamma\gamma) =
1.6\pm0.3)$\,\cite{ATLAS-CONF-2013-034}. The same can be said for the
signal strengths of $h_d$ and $h_s$ in the $\tau^+\tau^-$ channel. For
smaller non-zero values of \phkap\ $R_{h_d}(\tau^+\tau^-)$
($R_{h_s}(\tau^+\tau^-)$) is still large (small)
enough to be consistent with (excluded by) the LHC data, taking into
account the experimental errors on the measurements.  
Nevertheless, of the three cases discussed here, while this case
presents the possibility of the largest
contribution by $h_p$ to $h_d$ production at the LHC, it is the weakest in that
the signal strengths of the Higgs bosons predicted by it lie at the
verge of being excluded.  

\section{\label{sec:summary}Summary and outlook}

In this article we have presented the one-loop
Higgs mass matrix of the complex NMSSM in the RG-improved effective potential
approach, along with the expressions for Higgs boson trilinear
self-couplings. We have then
highlighted a scenario, precluded in the MSSM, wherein the decay of a
pseudoscalar-like Higgs boson into 125\,GeV Higgs bosons is induced
by non-zero values of the CPV phase \phkap. We have noted that, when
one of the scalar Higgs bosons is required to have
a SM-like signal rate, it is relatively easy for the mass of the
singlet-pseudoscalar-like Higgs boson to be near $\sim$
250\,GeV compared to the other heavy Higgs bosons of the model.
The fact that the decay width of a heavy Higgs boson into two lighter
ones is inversely proportional to its mass renders such a $\sim$
250\,GeV Higgs boson particularly interesting as well as relevant for
the phenomenology of the SM-like Higgs boson in the model.

We have analysed three benchmark cases corresponding to
different parameter configurations in the NMSSM which generate a
$\sim 125$\,GeV SM-like Higgs boson and a pseudoscalar
near 250\,GeV. In our analysis the impact of non-zero
CPV phases in each of these cases is quantified in terms of an
auxiliary signal rate $A^{h_p}(\gamma\gamma)$. 
This approximate quantity assumes that the $\sim
250$\,GeV pseudoscalar-like Higgs boson is produced in the
gluon fusion mode at the LHC and decays into a (pair of) SM-like Higgs
boson(s), one of which subsequently decays into a photon pair.
By calculating this auxiliary rate in each case studied, 
we have deduced that such a $\sim
250$\,GeV Higgs boson can generally contribute significantly to the production
of SM-like Higgs bosons at the LHC for large CPV phases. 
In fact, in one of the cases discussed, the auxiliary signal rate for this Higgs boson
can be as high as 25\% of the observed $\gamma\gamma$ rate. 

Evidently, a calculation of the total
cross section for our considered process is essential to draw concrete
inferences about its observability or significance at the LHC. 
In this regard, a calculation of higher order corrections to the Higgs trilinear
couplings in the complex NMSSM, following
those derived in \cite{Nhung:2013lpa} for the real NMSSM, could
prove crucial. Furthermore, a detailed study of the signal
topologies in various channels due to the production of multiple Higgs bosons, 
in line with the ones studied recently in \cite{Kang:2013rj,Ellwanger:2013ova,Musolf:2013xx}, 
is also in order. For this purpose, we eventually aim to embed the
cNMSSM in a publicly available tool such as CalcHEP\,\cite{Belyaev:2012qa} to make 
possible the calculation of actual cross sections in this model. 
Our current analysis, nevertheless, serves as a clear and timely
demonstration of the fact that CP violation
in the Higgs sector can be a very important probe of new physics at the LHC. 
Of particular relevance here is the observation that the $\sim
250$\,GeV Higgs boson mostly has a very poor signal strength when
decaying itself into a photon pair but a large BR into lighter Higgs
bosons for non-zero CPV phases. Thus, the already observed 
SM-like Higgs boson could provide an important, and possibly the 
only, handle on such a beyond-the-SM (and MSSM) scenario.  

\section*{Acknowledgements} 

The author is thankful to S.~Moretti for his valuable comments and suggestions for the
improvement of this draft. SM is funded in part by the Welcome 
Programme of the Foundation for Polish Science. 

\appendix
\numberwithin{equation}{section}

\section{Sparticle mass matrices}
\label{sec:app-A}

$\bullet$ The chargino mass matrix, in the
$(\widetilde{W}^-,\widetilde{H}^-)$ basis, using the convention
$\widetilde{H}^{-}_{L(R)} = \widetilde{H}^{-}_{d(u)}$, can be written as
\begin{eqnarray}
{\cal M}_C = \left(\begin{array}{cc}
     M_2              & \sqrt{2} M_W \cos\beta \\[2mm]
\sqrt{2} M_W \sin\beta &
\frac{|\lambda| v_S}{\sqrt{2}}\,e^{i\phlam'}
             \end{array}\right)\, ,
\end{eqnarray}
which is diagonalised by two different unitary matrices as
$ C_R {\cal M}_C C_L^\dagger ={\sf diag}\{m_{\widetilde{\chi}^\pm_1},\,
m_{\widetilde{\chi}^\pm_2}\}$, where
$m_{\widetilde{\chi}^\pm_1} \leq m_{\widetilde{\chi}^\pm_2}$. \\

\noindent$\bullet$ The neutralino mass matrix, in the
$(\widetilde{B},\widetilde{W}^0,\,\widetilde{H}^0_d,\,\widetilde{H}^0_u,\widetilde{S})$ basis, can be written as
\begin{eqnarray}
{\cal M}_N=\left(\begin{array}{ccccc}
  M_1       &      0          &  -m_Z \cos\beta s_W  & m_Z \sin\beta s_W  & 0\\[2mm]
0 &     M_2         &   m_Z \cos\beta c_W  & -m_Z \sin\beta c_W & 0\\[2mm]
 -m_Z \cos\beta s_W   &  m_Z \cos\beta c_W & 0 &
-\frac{|\lambda| v_S}{\sqrt{2}}\,e^{i\phlam'} &
-\frac{|\lambda| v s_\beta}{\sqrt{2}}\,e^{i\phlam'} \\[2mm]
m_Z \sin\beta s_W   & -m_Z \sin\beta c_W &  -\frac{|\lambda| v_S}{\sqrt{2}}\,e^{i\phlam'} &       0          &
-\frac{|\lambda| v \cos\beta}{\sqrt{2}}\,e^{i\phlam'}\\[2mm]
0 & 0 & -\frac{|\lambda| v s_\beta}{\sqrt{2}}\,e^{i\phlam'} &
-\frac{|\lambda| v \cos\beta}{\sqrt{2}}\,e^{i\phlam'} &
\sqrt{2} |\kappa| v_S \,e^{i\phkap'}
                  \end{array}\right)\,,
\end{eqnarray}
\noindent where $s_W = \sin\theta_W$, with $\theta_W$ being the
Weinberg angle. The above matrix is diagonalised as
$N^* {\cal M}_N N^\dagger = {\sf diag}\,
(m_{\widetilde{\chi}_1^0},m_{\widetilde{\chi}_2^0},
m_{\widetilde{\chi}_3^0},m_{\widetilde{\chi}_4^0},m_{\widetilde{\chi}_5^0})$,
where $N$ is a unitary matrix and
$m_{\widetilde{\chi}_1^0} \leq m_{\widetilde{\chi}_2^0} \leq m_{\widetilde{\chi}_3^0}
\leq m_{\widetilde{\chi}_4^0} \leq m_{\widetilde{\chi}_5^0}$. \\

\noindent$\bullet$ For the stop, sbottom and stau matrices, in the $\left(\widetilde{q}_L,
\widetilde{q}_R\right)$ basis, we have
\begin{eqnarray}
\widetilde{\cal M}^2_t  = \left( \begin{array}{cc}
M^2_{\widetilde{Q}_3}\, +\, m^2_t\, +\, \cos 2\beta M^2_Z\, (
\frac{1}{2} - \frac{2}{3} s_W^2 ) &
\frac{h_t^* v_u}{\sqrt{2}} (|A_t| e^{-i(\theta+\phi_{A_t})} -
\frac{|\lambda| v_S}{\sqrt{2}}e^{i\phlam'} \cot\beta)\\
\frac{h_t v_u}{\sqrt{2}} (|A_t| e^{i(\theta+\phi_{A_t})} -
\frac{|\lambda| v_S}{\sqrt{2}} e^{-i\phlam'} \cot\beta)
& \hspace{-0.2cm}
M^2_{\widetilde{U}_3}\, +\, m^2_t\, +\, \cos 2\beta M^2_Z\, Q_t s^2_W
\end{array}\right)\,, \nonumber
\end{eqnarray}
\begin{eqnarray}
\widetilde{\cal M}^2_b  = \left( \begin{array}{cc}
M^2_{\widetilde{Q}_3}\, +\, m^2_b\, +\, \cos 2\beta M^2_Z\, (
-\frac{1}{2} + \frac{1}{3} s_W^2 ) &
\frac{h_b^* v_d}{\sqrt{2}}(|A_b|e^{-i\phi_{A_b}} -
\frac{|\lambda| v_S}{\sqrt{2}}e^{i\phlam'} \tan\beta )/\sqrt{2}\\
\frac{h_b v_d}{\sqrt{2}} (|A_b|e^{i\phi_{A_b}} -
\frac{|\lambda| v_S}{\sqrt{2}}e^{-i\phlam'} \tan\beta )/\sqrt{2}
& \hspace{-0.2cm}
M^2_{\widetilde{D}_3}\, +\, m^2_b\, +\, \cos 2\beta M^2_Z\, Q_b s^2_W
\end{array}\right)\,,  \nonumber
\end{eqnarray}
\begin{eqnarray}
\widetilde{\cal M}^2_\tau  = \left( \begin{array}{cc}
M^2_{\widetilde{L}_3}\, +\, m^2_\tau\, +\, \cos 2\beta M^2_Z\,
( s_W^2-1/2 ) &
\frac{h_\tau^* v_d}{\sqrt{2}} (|A_\tau|e^{-i\phi_{A_\tau}} -
\frac{|\lambda| v_S}{\sqrt{2}}e^{i\phlam'} \tan\beta )/\sqrt{2}\\
\frac{h_\tau v_d}{\sqrt{2}} (|A_\tau|e^{i\phi_{A_\tau}} -
\frac{|\lambda| v_S}{\sqrt{2}}e^{-i\phlam'} \tan\beta )/\sqrt{2}
& \hspace{-0.2cm}
M^2_{\widetilde{E}_3}\, +\, m^2_\tau\, -\, \cos 2\beta M^2_Z\, s^2_W
\end{array}\right)\,,\nn \\
\end{eqnarray}
\noindent where $h_\tau \equiv \frac{2 m_\tau}{v_d}$ and $m_\tau$ are
the Yukawa coupling and mass of the $\tau$ lepton, respectively, and
$A_\tau\equiv |A_\tau|e^{i\phi_{A_\tau}}$ is the soft Yukawa coupling
of $\tilde{\tau}$. The mass
eigenstates of the top and bottom squarks and the stau are obtained by diagonalising the above mass matrices 
as $U^{\tilde{f}\dagger} \, \widetilde{\cal M}^2_f \,
U^{\tilde{f}} ={\sf
  diag}(m_{\tilde{f}_1}^2,m_{\tilde{f}_2}^2)\,$, such that
$m_{\tilde{f}_1}^2 \leq m_{\tilde{f}_2}^2$, for $f=t, b$ and $\tau$.

\section{Functions}
\label{sec:app-B}

$\bullet$ The functions used in the leading (s)quark
corrections to the Higgs mass matrix are given as
\begin{eqnarray}
L_{\tilde{t}} &= &\ln{\left(\frac{m_{\tilde{t}_2}^2}{m_{\tilde{t}_1}^2}\right)}\;,\;\;\;
L_{\tilde{b}} = \ln{\left(\frac{m_{\tilde{b}_2}^2}{m_{\tilde{b}_1}^2}\right)}\,, \nn \\
L_{\tilde{t}t} &= &\ln{\left(\frac{m_{\tilde{t}_1} m_{\tilde{t}_2}}{m_t^2}\right)}\;,\;\;\;
L_{\tilde{b}b} = \ln{\left(\frac{m_{\tilde{b}_1} m_{\tilde{b}_2}}{m_b^2}\right)}\,, \nn \\
f_t &=& \frac{1}{m_{\tilde{t}_2}^2-m_{\tilde{t}_1}^2}\Big[
m_{\tilde{t}_2}^2
\ln{\left(\frac{m_{\tilde{t}_2}^2}{M_{\rm SUSY}^2}\right)} - 
m_{\tilde{t}_1}^2\ln{\left(\frac{m_{\tilde{t}_1}^2}{M_{\rm SUSY}^2}
\right)} \Big]-1\,,\nn \\
f_b &=& \frac{1}{m_{\tilde{b}_2}^2-m_{\tilde{b}_1}^2} \Big[
m_{\tilde{b}_2}^2
\ln{\left(\frac{m_{\tilde{b}_2}^2}{M_{\rm SUSY}^2}\right)} - 
m_{\tilde{b}_1}^2\ln{\left(\frac{m_{\tilde{b}_1}^2}{M_{\rm SUSY}^2}
\right)} \Big]-1\,,\nn \\
g_t &=& \left[\frac{m_{\tilde{t}_2}^2 + m_{\tilde{t}_1}^2}{m_{\tilde{t}_2}^2 -
m_{\tilde{t}_1}^2} L_{\tilde{t}}  -2\right]\;,\;\;\;
g_b = \left[\frac{m_{\tilde{b}_2}^2 + m_{\tilde{b}_1}^2}{m_{\tilde{b}_2}^2 -
m_{\tilde{b}_1}^2} L_{\tilde{b}} -2\right]\,,
\label{c.16e}
\end{eqnarray}
where the mass eigenvalues $m_{\tilde q}$ have been given in Appendix\,\ref{sec:app-A}. \\

\noindent$\bullet$ Additional quantities used in the $D$-term contributions
are given as
\begin{eqnarray}
g_u &=& \frac{1}{4} g^2_2 - \frac{5}{12}g_1^2 \;,\;\;\;
g_d = \frac{1}{4} g^2_2 - \frac{1}{12}g_1^2 \,,\nonumber \\
D_u &=&
\frac{1}{2}\Big(M_{\tilde{Q}_3}-M_{\tilde{U}_3}+\frac{g_u}{2}(v_d^2-v_u^2)\Big) \,,
\nonumber \\
D_d &=& \frac{1}{2}\Big(M_{\tilde{Q}_3}-M_{\tilde{D}_3}+\frac{g_d}{2}(v_u^2-v_d^2)\Big) \,,
\nonumber \\
C_t &=& \frac{3 m_t^2 }{32\pi^2} \left[\frac{4g_u D_u}{(m_{\tilde{t}_2}^2 -
  m_{\tilde{t}_1}^2)^2}g'_t - \frac{g_1^2 + g_2^2}{2(m_{\tilde{t}_2}^2 -
  m_{\tilde{t}_1}^2)}L_{\tilde{t}} \right] \,,\nonumber \\
C_b &=& \frac{3 m^2_b}{32\pi^2} \left[\frac{4g_d D_d }{(m_{\tilde{b}_2}^2 -
  m_{\tilde{b}_1}^2)^2}g'_b - \frac{g_1^2 + g_2^2}{2(m_{\tilde{b}_2}^2 -
  m_{\tilde{b}_1}^2)}L_{\tilde{b}} \right] \,,\nonumber \\
D_t &=& - \frac{3m^2_t}{16\pi^2}\left[ \frac{2g_uD_u}{(m_{\tilde{t}_2}^2 -
  m_{\tilde{t}_1}^2)}L_{\tilde{t}} 
+ \frac{g_1^2 + g_2^2}{4}\ln\left(\frac{m_{\tilde{t}_1}^2
  m_{\tilde{t}_2}^2}{M^2_{\rm SUSY}}\right) \right] \,,\nonumber 
\end{eqnarray}
\begin{eqnarray}
D_b &=& - \frac{3m^2_b}{16\pi^2}\left[ \frac{2g_dD_d}{(m_{\tilde{b}_2}^2 -
  m_{\tilde{b}_1}^2)}L_{\tilde{b}} 
+ \frac{g_1^2 + g_2^2}{4}\ln\left(\frac{m_{\tilde{b}_1}^2
  m_{\tilde{b}_2}^2}{M^2_{\rm SUSY}}\right) \right] \,.
\end{eqnarray}

\def\lm2mu{\ln\left(\frac{\operatorname{max}
(M_{1,2}^2,|\mu|^2)}{M_\mathrm{SUSY}^2}\right)}
\def\lmunu{\ln\left(\frac{\operatorname{max}
(4|\nu|^2,|\mu|^2)}{M_\mathrm{SUSY}^2}\right)}
\def\lnu{\ln\left(\frac{4|\nu|^2}{M_\mathrm{SUSY}^2}\right)}
\def\lmu{\ln\left(\frac{|\mu|^2}{M_\mathrm{SUSY}^2}\right)}

\noindent$\bullet$ The chargino/neutralino corrections use the following potentially large logarithms:
\begin{eqnarray}
L_{\mu} &=& \lmu\;,\;\;\;
L_{\nu} = \lnu\,, \nn \\
L_{M_2\mu} &=& \lm2mu\;,\;\;\;
L_{\mu\nu} = \lmunu\,,
\label{c.22e} 
\end{eqnarray}
where for simplification we assume $M_1 \sim M_2
\equiv M_{1,2}$ for the gaugino masses. \\

\noindent $\bullet$ The Higgs wave function renormalisation constants for the
three weak eigenstates $H_u$, $H_d$ and $S$ are given, in the Landau gauge, as 
\begin{eqnarray}
Z_{H_u} &=& 1+\frac{1}{16\pi^2}\Big[
3h_t^2 \ln\left(\frac{M_{\rm SUSY}^2}{m_{t}^2}\right)
-\frac{3}{4}(g_1^2+3g_2^2) \ln\left(\frac{M_{\rm SUSY}^2}{m_Z^2}\right)
\nn \\
&&+\cos^2\b (3h_b^2 +h_\tau^2 -3h_t^2) \ln\left(\frac{M_A^2}{m_{t}^2}\right)
+\frac{g_1^2}{2} \ln\left(\frac{M_{\rm SUSY}^2}{\operatorname{max}(|\mu|^2,M_1^2)}\right)\nn \\
&&+\frac{3g_2^2}{2}
\ln \left(\frac{M_{\rm SUSY}^2}{\operatorname{max}(|\mu|^2,M_2^2)}\right)
+\lambda^2
\ln \left(\frac{M_{\rm SUSY}^2}{\operatorname{max}(|\mu|^2,4|\nu|^2)}\right)
\Big]\,, \nn \\
Z_{H_d} &=& 1+\frac{1}{16\pi^2}\Big[
(3h_b^2+h_\tau^2) \ln \left(\frac{M_{\rm SUSY}^2}{m_{t}^2}\right)
-\frac{3}{4}(g_1^2+3g_2^2)
\ln \left(\frac{M_{\rm SUSY}^2}{m_Z^2}\right)
\nn \\
&&+\sin^2\b (3h_t^2 -h_\tau^2 -3h_b^2) \ln \left(\frac{M_A^2}{m_{t}^2}\right)
+\frac{g_1^2}{2}
\ln \left(\frac{M_{\rm SUSY}^2}{\operatorname{max}(|\mu|^2,M_1^2)}\right)\nn
\\
&&+\frac{3g_2^2}{2}
\ln \left(\frac{M_{\rm SUSY}^2}{\operatorname{max}(|\mu|^2,M_2^2)}\right)
+\l^2
\ln \left(\frac{M_{\rm SUSY}^2}{\operatorname{max}(|\mu|^2,4|\nu|^2)}\right)
\Big]\,,\nn \\
Z_S &=& 1+\frac{1}{8\pi^2}\Big[\lambda^2
\ln \left(\frac{M_{\rm SUSY}^2}{|\mu|^2}\right) +\kappa^2
\ln \left(\frac{M_{\rm SUSY}^2}{4|\nu|^2}\right)\Big]\,.
\label{c.11e}
\end{eqnarray}

\bibliographystyle{utphysmcite}
\bibliography{CPV-refs}

\end{document}